\documentclass[fleqn,usenatbib]{mnras}

\usepackage{newtxtext,newtxmath}
\usepackage[T1]{fontenc}

\DeclareRobustCommand{\VAN}[3]{#2}
\let\VANthebibliography\thebibliography
\def\thebibliography{\DeclareRobustCommand{\VAN}[3]{##3}\VANthebibliography}

\usepackage{graphicx}	% Including figure files
\usepackage{amsmath}	% Advanced maths commands
\usepackage{ulem}

\newcommand{\dd}{{\rm d}}
\newcommand{\lp}{\left(}
\newcommand{\rp}{\right)}
\newcommand{\msunyr}{\ensuremath{\rm{M}_\odot\,\text{yr}^{-1}}}
\newcommand{\ctgas}{\ensuremath{c_{T,\text{gas}}}}
\newcommand{\vesc}{\ensuremath{v_{\text{esc}}}}
\newcommand{\vcrit}{\ensuremath{v_\text{crit}}}
\newcommand{\vout}{\ensuremath{v_\text{out}}}
\newcommand{\msun}{\ensuremath{\rm{M}_\odot}}
\newcommand{\deltaRL}{\ensuremath{\delta R_\text{L}}}
\newcommand{\mdot}{\ensuremath{\dot{M}}}
\newcommand{\mdotoutLone}{\ensuremath{\mdot_\text{out,L1}}}
\newcommand{\mdotd}{\ensuremath{\mdot_\text{d}}}

\defcitealias{cehula2023}{CP23}

\title[Effects of radiation on binary mass transfer]{On the effects of radiation on mass transfer in binary stars}

\author[Cehula \& Pejcha]{
Jakub Cehula$^{1,2,3}$\thanks{E-mail: jakub.cehula@ut.ee} and
Ond\v{r}ej Pejcha$^{1}$
\\
$^{1}$Institute of Theoretical Physics, Faculty of Mathematics and Physics, Charles University, V Hole\v{s}ovi\v{c}k\'{a}ch 2, Prague, 180 00, Czech Republic\\
$^{2}$Tartu Observatory, University of Tartu, T\~{o}ravere, 61602 Tartumaa, Estonia \\
$^{3}$Astronomical Institute, Slovak Academy of Sciences, 059 60 Tatransk\'{a} Lomnica, Slovakia
}

\date{Accepted 2025 December 18. Received 2025 November 19; in original form 2025 May 13}

\pubyear{\the\year{}}

\begin{document}
\label{firstpage}
\pagerange{\pageref{firstpage}--\pageref{lastpage}}
\maketitle

\begin{abstract}
Mass transfer (MT) in binary systems is a common evolutionary process that can significantly affect the structure, evolution, and final fate of both stars. In modeling MT hydrodynamics, it is usually assumed that the critical point of the flow, where the velocity exceeds the local sound speed, coincides with the inner Lagrange point (L1). However, in massive donors where radiative pressure dominates over gas pressure and the Eddington factor $\Gamma_\text{Edd}$ can approach or exceed unity, radiation–gas coupling can shift the critical point away from L1, altering the MT rate ($\mdotd$). We investigate the effects of radiation on MT using time-steady radiative hydrodynamic equations and the von Zeipel theorem. We derive analytical expressions that closely approximate $\mdotd$, algebraic solutions for simplified cases, and numerical results using a realistic equation of state. Two main differences emerge relative to traditional prescriptions for $\mdotd$. First, for Roche-lobe-underfilling donors with $\Gamma_\text{Edd} \lesssim 1$, radiative momentum exchange leads to an exponential increase of $\mdotd$ as a function of $1-\Gamma_\text{Edd}$. We provide a simple modification of existing prescriptions that captures this effect. Second, the photon tiring limit for super-Eddington outflows is much less restrictive near L1 than in spherical stars. We suggest that donors with super-Eddington, convectively inefficient subsurface layers can drive MT with $-\mdotd \gtrsim 10^{-2}\,\msunyr$ even before Roche-lobe overflow. We characterize the conditions for this new mode of super-Eddington-boosted MT and discuss its implications for binary evolution, including potential links to nonterminal outbursts of Luminous Blue Variables.
\end{abstract}

\begin{keywords}
binaries: close -- stars: evolution -- stars: mass-loss --  radiation: dynamics
\end{keywords}

%%%%%%%%%%%%%%%%%%%%%%%%%%%%%%%%%%%%%%%%%%%%%%%%%%

%%%%%%%%%%%%%%%%% BODY OF PAPER %%%%%%%%%%%%%%%%%%

\section{Introduction}

Massive stars drive cosmic evolution through their winds, photons, and explosions. Their evolutionary journey ultimately leads to the synthesis of new chemical elements and the formation of neutron stars and black holes. Most massive stars exist in gravitationally bound binaries or higher-order multiples \citep[e.g.][]{moe2017} and mutual interaction between stars can significantly alter their evolution, ultimate fate, and imprint on surroundings \citep[e.g.][]{sana2012,tauris2023,marchant2024}. Binary stars often interact through mass transfer (MT), where material flows from one star through the first Lagrange point (L1) to its companion. Ongoing MT can be detected by observing the energy released as the material falls toward the companion or by identifying signatures of diffuse gas surrounding the binary system when some material escapes the binary completely. Over time, MT can significantly alter both the internal structures and external appearances of stars, making evolutionary paths impossible for isolated stars. In some cases, MT becomes unstable, triggering common envelope evolution, which is a critical phase in many evolutionary pathways that lead to compact object binaries and gravitational wave progenitors \citep{paczynski1976,belczynski2002,mandel2022,roekpke2023}.

One key difference between massive stars and low-mass stars is the significant role of radiation pressure. The increasing importance of radiation pressure with stellar mass, both inside and outside the photosphere, manifests in relations between global properties of stars, such as the mass--luminosity relation on the main sequence, the intensification of line-driven winds, and positively contributes to the viability of phenomena such as chemically homogeneous evolution \citep[e.g.,][]{maeder_meynet87,maeder87,heger00,vink00,yoon06}.

Massive stars can exceed the continuum Eddington limit, where the outward frequency-integrated radiative acceleration exceeds the inward gravitational pull. The existence and position of super-Eddington regions depend on density, temperature, and opacity profiles inside the star. Historically, approaching or exceeding the Eddington limit has been argued to be associated with increased mass loss. There is observational evidence that the proximity to the Eddington limit is the physical reason for the onset of Wolf-Rayet-type mass loss \citep[e.g.][]{grafener2011} and a potential mechanism responsible for luminous blue variable (LBV) eruptions \citep[e.g.,][]{lamers88,smith_owocki06,smith2011}. 

The simple fact of existence of a super-Eddington region is not sufficient to generate outflows because the energy flux can instead be carried out by convection \citep{joss73}. Near the stellar surface, convection can become inefficient once the required convective velocities approach or exceed the local sound speed. One-dimensional stellar models are able to support limited convectively inefficient super-Eddington regions by means of envelope inflation, density and pressure inversions, and possibly pulsations \citep[e.g.,][]{glatzel93,sanyal2015}. \citet{owocki2004} argued that the mass-loss rate of any super-Eddington wind driven from the region of inefficient convection would far exceed the photon tiring limit, where all available stellar luminosity is used to lift material from the star to infinity, and such outflow would stagnate at some finite radius and fall back to the star. Consequently, multidimensional effects such as hydrodynamic instabilities and effectively porous envelopes for radiation transfer are likely to significantly alter these spherically symmetric predictions \citep[e.g.,][]{mihalas69,shaviv01a,begelman01,owocki2004}. This was confirmed by time-dependent simulations of \citet{jiang2015,jiang2018}, which showed complex outflow and inflow patterns with occasional eruptions. Nevertheless, continuum super-Eddington winds can still occur if some additional energy is deposited below the surface, and the photon-tiring limit remains important in these situations \citep{shaviv01b,owocki2004,owocki2017,quataert16,shen16}.

The ideas and results on super-Eddington regions in stars rely on the crucial fact that radiative and gravitational accelerations in a spherically symmetric object have the same inverse-square scaling with radius. This is not the case for Roche-lobe-filling stars in binaries, where effective gravity vanishes at the L1 point. Consequently, in a fraction of the stellar envelope, radiative forces can potentially overwhelm gravity and drive much higher MT rates because the photon tiring now needs to overcome much smaller gravitational potential difference. Such a super-Eddington MT boost could play an important role in binary evolution involving massive stars and for explaining the origin of eruptions in LBVs, which are or could have once been a component of a binary system \citep[e.g.,][]{smith_owocki06,smith2015,mahy2022}.

\begin{figure}
    \centering
    \includegraphics[width=\columnwidth]{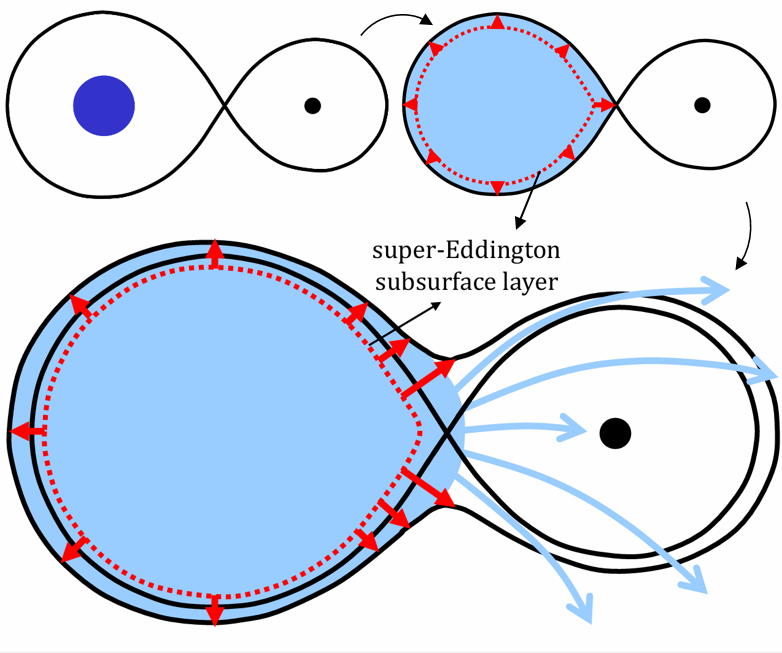}
    \caption{Cartoon illustrating the idea of the super-Eddington mass-transfer boost.}
    \label{fig:super-Edd}
\end{figure}

In this paper, our objective is to characterize the effects of radiation pressure on MT through the L1 point. In particular, we are interested in understanding MT from donors with a super-Eddington layer located immediately inside the L1 point, as illustrated in Fig.~\ref{fig:super-Edd}. Our approach builds on our previous work in \citet[hereafter \citetalias{cehula2023}]{cehula2023}, where we averaged the three-dimensional equations of hydrodynamics in the plane normal to the axis connecting both stars and solved the resulting one-dimensional two-point boundary value problem. In other words, we solved the structure of the star just interior the L1 point allowing for velocity terms and connecting to the hydrostatic spherically symmetric stellar model at the inner boundary. Here, we modify the method to include radiation. By finding a steady hydrodynamic solution in a gravitational potential near L1, we are in principle able to isolate effects that are inaccessible in the traditional approach, where the MT rate is calculated by post-processing a spherically symmetric stellar model \citep[e.g.][]{ritter1988,kolb1990,jackson2017,marchant2021}.

In Section~\ref{sec:theory}, we introduce the general 3D and 1D time-steady radiation hydrodynamic equations in flux-limited diffusion together with the von Zeipel theorem. In Section~\ref{sec:limiting}, we simplify the equations for the limiting cases of hydrostatic equilibrium, small velocities, and adiabaticity. In Section~\ref{sec:subedd}, we investigate sub-Eddington donors by constructing analytical estimates and presenting algebraic and numerical solutions. We also illustrate our results on realistic massive donors. In Section~\ref{sec:superedd}, we develop the theory of super-Eddington MT boost and apply the concept to a previously calculated binary evolution trajectory. In Section~\ref{sec:discussion}, we summarize our findings and discuss the implications and directions of future research.

\section{Theory}\label{sec:theory}

Here, we introduce the radiation hydrodynamic equations that govern MT. In Section~\ref{sec:3D}, we use general 3D time-steady equations in the flux-limited diffusion approximation to derive a general velocity (momentum) equation. In Section~\ref{sec:1D}, we reduce the equations to 1D assuming that the flow proceeds through a nozzle around the L1 point. We also compare this case with a spherically symmetric stellar wind. Finally, we introduce the von Zeipel theorem in Section~\ref{sec:von_zeipel}. This is our crucial assumption, which allows us to reduce the equations to the limiting cases in Section~\ref{sec:limiting}, and which we further solve in Section~\ref{sec:subedd}.

\subsection{General three-dimensional equations}\label{sec:3D}
We start with stationary frequency-integrated 3D radiation hydrodynamic equations assuming that radiation has a blackbody spectrum at the local radiation temperature, scattering is negligible, flux mean opacity is equal to Rosseland mean opacity (which is valid in optically thick regions), and using flux-limited diffusion approximation \citep[e.g.][]{alme1973,mihalas1984,krumholz2007}.  The mixed-frame formulation correct to $\mathcal{O} (v/c)$ in the streaming, static diffusion, and dynamic diffusion regimes including the Roche potential $\phi_{\rm R}$ reads \citep[e.g.][]{zhang2011,calderon2021}
\begin{subequations}\label{eq:3D}
	\begin{align}
	   & \boldsymbol{\nabla} \boldsymbol{\cdot} \lp\rho \boldsymbol{v}\rp = 0 , \label{eq:3D_1}\\
	     & \boldsymbol{\nabla} \boldsymbol{\cdot} \lp \rho \boldsymbol{v} \otimes \boldsymbol{v} \rp + \boldsymbol{\nabla} P_{\rm gas} + \lambda \boldsymbol{\nabla} E_{\rm rad} = -\rho \boldsymbol{\nabla} \phi_{\rm R}, \label{eq:3D_2}\\
	     & \boldsymbol{\nabla} \boldsymbol{\cdot} \lp \rho \epsilon^* \boldsymbol{v} + P_{\rm gas} \boldsymbol{v} \rp + \lambda \boldsymbol{v} \cdot \boldsymbol{\nabla} E_{\rm rad} = - c \rho \kappa_{\rm P} \lp a_{\rm rad}T^4 - E_{\rm rad}^{(0)} \rp, \label{eq:3D_3}\\
          & \boldsymbol{\nabla} \boldsymbol{\cdot} \lp \frac{3-f}{2} E_{\rm rad} \boldsymbol{v}\rp - \lambda \boldsymbol{v} \cdot \boldsymbol{\nabla} E_{\rm rad} = c \rho \kappa_{\rm P} \lp a_{\rm rad}T^4 - E_{\rm rad}^{(0)} \rp \nonumber \\
          & \quad + \boldsymbol{\nabla} \cdot \lp \frac{c \lambda}{\rho \kappa_{\rm R}} \boldsymbol{\nabla} E_{\rm rad}\rp, \label{eq:3D_4}\\
          & \boldsymbol{F}_{\rm rad}^{(0)} = - \frac{c \lambda}{\rho \kappa_{\rm R}} \boldsymbol{\nabla} E_{\rm rad}^{(0)}, \label{eq:3D_5}\\
          & P_{\rm rad}^{(0)} = f E_{\rm rad}^{(0)}, \label{eq:3D_6}
        \end{align}
\end{subequations}
where the quantities with the superscript $(0)$ are evaluated in the co-moving frame while quantities without it in the lab frame.  Gas density, thermal pressure, and velocity are denoted by $\rho$, $P_{\rm gas}$, and $\boldsymbol{v}$, respectively.  Radiation energy density, pressure, and flux are denoted by $E_{\rm rad}$, $P_{\rm rad}$, and $\boldsymbol{F}_{\rm rad}$, respectively.  Next, $c$ is the speed of light, $a_{\rm rad}$ the radiation constant, $T$ the gas temperature, and $\kappa_{\rm P}$ and $\kappa_{\rm R}$, the Planck mean and Rosseland mean opacities, respectively.  It is worth noting that the Planck mean and Rosseland mean interaction coefficients $\rho \kappa_{\rm P}$ and $\rho \kappa_{\rm R}$ (in units of cm$^{-1}$) are always evaluated in the co-moving frame.  The total gas specific energy $\epsilon^*$ comprises of three components, internal, kinetic, and potential,
\begin{equation}
    \epsilon^* = \epsilon_{\rm gas} + \frac{1}{2} \left|\boldsymbol{v}\right|^2 + \phi_{\rm R},\label{eq:epsilon_star}
\end{equation}
where $\epsilon_{\rm gas}$ is the gas (internal) specific energy in units of $\text{erg} \: \text{g}^{-1}$.  The flux limiter $\lambda$ and the Eddington factor $f$ are defined by \citep{levermore1981}
\begin{equation}\label{eq:flux_lim}
    \lambda = \frac{2+R}{6+3R+R^2}, \quad f = \lambda + \lambda^2 R^2, \quad R = \frac{\left| \boldsymbol{\nabla} E_{\rm rad}^{(0)}\right|}{\rho \kappa_{\rm R} E_{\rm rad}^{(0)}}.
\end{equation}
Radiation energy density in the lab frame and the co-moving frame are related by \citep{zhang2011}
\begin{equation}
    E_{\rm rad}^{(0)} = E_{\rm rad} - \frac{2}{c^2} \boldsymbol{v} \cdot \boldsymbol{F}_{\rm rad}^{(0)} + \mathcal{O} \lp \frac{v^2}{c^2}\rp = E_{\rm rad} + 2 \frac{\lambda}{\rho \kappa_{\rm R}} \frac{\boldsymbol{v}}{c} \cdot \boldsymbol{\nabla} E_{\rm rad} + \mathcal{O} \lp \frac{v^2}{c^2}\rp.
\end{equation}
In some parts of our work we further assume ideal gas equation of state (EOS),
\begin{equation}\label{eq:id}
    P_{\rm gas} = \frac{k}{\overline{m}} T \rho = c_{T, \rm gas}^2 \rho, \quad \epsilon_{\rm gas} = \frac{1}{\gamma-1} \frac{k}{\overline{m}} T = \frac{1}{\gamma-1} c_{T, \rm gas}^2,
\end{equation}
where $k$ is the Boltzmann constant, $\overline{m}$ is the mean mass of a gas particle, $c_{T, \rm gas}$ is the gas isothermal sound speed, and $\gamma = c_P / c_V$, where $c_P$ and $c_V$ are dimensionless heat capacities at constant pressure and volume.  

Based on these assumptions, we can derive a velocity equation for ideal gas in the form following \citet[eq.~10]{lubow1975}, which elucidates the terms contributing to the existence and position of the critical point,
\begin{equation}\label{eq:3D_vel_id}
    \boldsymbol{\nabla} \boldsymbol{\cdot} \boldsymbol{v} = \frac{\left| \boldsymbol{v} \right|^3 \boldsymbol{\nabla} \boldsymbol{\cdot} \lp {\boldsymbol{v}}/{\left| \boldsymbol{v}\right|}\rp - \lp \gamma-1 \rp \left| \boldsymbol{v} \right| q_{\rm rad} + \boldsymbol{v} \boldsymbol{\cdot} \boldsymbol{f}_{\rm rad} - \boldsymbol{v} \boldsymbol{\cdot} \boldsymbol{\nabla} \phi_{\rm R}}{\left|\boldsymbol{v}\right|^2 - c_{s, \rm gas}^2},
\end{equation}
where we denote
\begin{equation}\label{eq:q_rad}
    q_{\rm rad} \equiv - \frac{c}{\left| \boldsymbol{v}\right|} \kappa_{\rm P} \lp a_{\rm rad}T^4 - E_{\rm rad}^{(0)}\rp,
\end{equation}
\begin{equation}\label{eq:f_rad}
    \boldsymbol{f}_{\rm rad} \equiv - \frac{\lambda}{\rho} \boldsymbol{\nabla} E_{\rm rad} = \frac{\kappa_{\rm R}}{c} \lp \boldsymbol{F}_{\rm rad} - \boldsymbol{F}_{\rm adv}\rp,
\end{equation}
and where we subtract the advective flux $\boldsymbol{F}_{\rm adv} \equiv \boldsymbol{v} E_{\rm rad} + \boldsymbol{v \cdot} \boldsymbol{\mathsf{P}}_{\rm rad}$ from the lab frame radiative flux $\boldsymbol{F}_{\rm rad}$, where $\boldsymbol{\mathsf{P}}_{\rm rad}$ is the radiation pressure tensor \citep{krumholz2007}. The gas adiabatic sound speed $c_{s, \rm gas}$ reads
\begin{equation}\label{eq:c_s_gas}
    c_{s, \rm gas}^2 = \gamma \frac{P_{\rm gas}}{\rho}.
\end{equation}
The velocity equation for a general EOS (\ref{eq:3D_vel}) together with its derivation can be found in Appendix \ref{app:3D}. This more complete equation differs only by additional terms arising from the thermodynamic derivatives of the EOS, which do not fundamentally change the structure.

There are three differences between our velocity equation~(\ref{eq:3D_vel_id}) and the isothermal velocity equation of \citet[][their eq. 10]{lubow1975}.  First, our velocity equation includes the term $\boldsymbol{v \cdot} \boldsymbol{f}_{\rm rad}$ that accounts for the increase in radiation momentum originating from the radiation pressure.  Second, our equation involves the term $- \lp \gamma-1\rp \left| \boldsymbol{v}\right| q_{\rm rad}$ that accounts for the gas specific heating/cooling rate (in units of $\text{erg}\:\text{g}^{-1}\:\text{s}^{-1}$) due to energy exchange between the gas and radiation. Third, our critical velocity is the gas adiabatic sound speed $c_{s,\rm gas}$ instead of the gas isothermal sound speed $c_{T, \rm gas}$.  

The sonic surface ($|v|=c_{s,\text{gas}}$)  is located at a region where the right-hand-side numerator of equation~(\ref{eq:3D_vel_id}) vanishes,
\begin{equation}
    \left| \boldsymbol{v}\right|^2 \boldsymbol{\nabla \cdot} \lp \frac{\boldsymbol{v}}{\left| \boldsymbol{v}\right|}\rp = \lp \gamma -1 \rp q_{\rm rad} + \frac{\boldsymbol{v}}{\left| \boldsymbol{v}\right|} \boldsymbol{\cdot} \lp \boldsymbol{\nabla} \phi_{\rm R} - \boldsymbol{f}_{\rm rad}\rp.
\end{equation}
Thus, heating/cooling due to energy exchange between gas and radiation $q_{\rm rad}$ and radiation momentum boost $\boldsymbol{f}_{\rm rad}$ can move the location of the sonic surface away from the L1 point ($\boldsymbol{\nabla} \phi_{\rm R} = 0$).

\subsection{General one-dimensional equations}\label{sec:1D}
Let us now assume that the gas flows through a nozzle oriented along the $x$ axis and with a cross-section denoted by $Q(x)$. In \citetalias{cehula2023}, we calculated $Q(x)$ based on the Roche potential and thermodynamic properties of the gas flowing to the L1 point. Assuming that all quantities are constant at a given $Q(x)$, we can reduce the set of 3D equations~(\ref{eq:3D}) to the set of 1D equations (see Appendix~\ref{app:1D} for details),
\begin{subequations}\label{eq:1D}
    \begin{align}
        & \frac{1}{v}\frac{\dd v}{\dd x} + \frac{1}{\rho}\frac{\dd \rho}{\dd x} + \frac{1}{Q}\frac{\dd Q}{\dd x} = 0, \label{eq:1D_mass}\\
        & v \frac{\dd v}{\dd x} + \frac{1}{\rho}\frac{\dd P_{\rm gas}}{\dd x} = f_{\rm rad} - \frac{\dd \phi_{\rm R}}{\dd x}, \label{eq:1D_mom}\\
        & \frac{\dd \epsilon_{\rm gas}}{\dd x} - \frac{P_{\rm gas}}{\rho} \frac{1}{\rho}\frac{\dd \rho}{\dd x} = q_{\rm rad}, \label{eq:1D_en}\\
        & L_{\rm N} = \left[ \frac{c}{\kappa_{\rm R}} f_{\rm rad} + \lp \rho \epsilon^* + P_{\rm gas} + \frac{3 - f}{2} E_{\rm rad} \rp v \right] Q, \label{eq:1D_Frad}
    \end{align}
\end{subequations}
where $L_{\rm N} = \rm{const.}$ is the total energy flux being transferred through the nozzle and comprises of the radiative $L_{\rm N, rad} = ({c}/{\kappa_{\rm R}}) f_{\rm rad} Q$ and advective $L_{\rm N, adv} = \lp \rho \epsilon^* + P_{\rm gas} + (3-f)E_{\rm rad}/2\rp v Q$ part. Instead of equation~(\ref{eq:3D_5}) we can use the 1D version of equation~(\ref{eq:f_rad}), i.e.
\begin{equation}\label{eq:f_rad_1D}
    f_{\rm rad} = - \frac{\lambda}{\rho} \frac{\dd E_{\rm rad}}{\dd x},
\end{equation}
and equation~(\ref{eq:3D_6}) remains unchanged. The donor's mass loss rate is given by \citepalias[see also][]{cehula2023}
\begin{equation}\label{eq:Mdot}
    - \dot{M}_{\rm d} = v \rho Q.
\end{equation}
Assuming ideal gas EOS (equation~\ref{eq:id}) we can derive the velocity equation in the following form
\begin{equation}\label{eq:1D_vel_id}
    \frac{1}{v}\frac{\dd v}{\dd x} = \frac{\displaystyle c_{s, \rm gas}^2 \frac{1}{Q} \frac{\dd Q}{\dd x} - \lp \gamma - 1 \rp  q_{\rm rad} + f_{\rm rad} - \frac{\dd \phi_{\rm R}}{\dd x}}{v^2 - c_{s, \rm gas}^2}.
\end{equation}
The velocity equation for a general EOS can be found in Appendix~\ref{app:1D}.

In a special case of spherically symmetric and isolated donor (star), i.e. spherically symmetric mass loss (nozzle), we can write $x \to r$, $Q \to 4 \pi r^2$, and $\phi_{\rm R} \to \phi_{\rm d} =  - G M_{\rm d}/r$, where $r$ is the radial coordinate, $\phi_{\rm d}$ is the donor's gravitational potential, $G$ is the gravitational constant, and $M_{\rm d}$ is donor's mass. This leads to the velocity equation in the form
\begin{equation}
    \frac{1}{v}\frac{\dd v}{\dd r} = \frac{\displaystyle c_{s, \rm gas}^2 \frac{2}{r} - \lp \gamma - 1 \rp  q_{\rm rad} + f_{\rm rad} - \frac{G M_{\rm d}}{r^2}}{v^2 - c_{s, \rm gas}^2},
\end{equation}
which is the general form of the momentum equation of a spherically symmetric stellar wind with energy input $\lp q_{\rm rad}\rp$ and momentum input $\lp f_{\rm rad}\rp$ \citep[eq.~4.18]{lamers1999}. Moreover, considering equation~(\ref{eq:1D_Frad}), we get $f_{\rm rad} = (L_{\rm d,rad}/L_{\rm d,Edd}) (G M_{\rm d}/r^2) =  (L_{\rm d,rad}/L_{\rm d,Edd}) (\dd \phi_{\rm R}/\dd r)$, where $L_{\rm d,rad}$ is the donor's radiative (diffusive) luminosity and $L_{\rm d,Edd}$ is the donor's Eddington luminosity and we have used $L_{\rm N} = L_{\rm d}$, where $L_{\rm d}$ is the donor's total luminosity and $L_{\rm d,Edd} = 4 \pi c G M_{\rm d}/\kappa_{\rm R}$.

\subsection{The von Zeipel theorem}\label{sec:von_zeipel}
The set of 1D radiation hydrodynamic equations~(\ref{eq:1D}) can be further simplified. We can use an extension of the von Zeipel theorem, which states that the radiative flux in a uniformly rotating star is proportional to the local effective gravity \citep{vonZeipel1924}, and write for the donor's envelope \citep[see also recent works by][]{fabry2022,fabry2023}
\begin{equation}\label{eq:von_Zeipel}
    \boldsymbol{F}_{\rm rad} = \frac{L_{\rm rad} (\phi_{\rm R})}{4 \pi G M (\phi_{\rm R})} \boldsymbol{\nabla} \phi_{\rm R},
\end{equation}
where $L_{\rm rad}(\phi_{\rm R})$ and $M(\phi_{\rm R})$ are the donor's radiative luminosity and mass coordinates, respectively, corresponding to a given equipotential $\phi_{\rm R}$. 

Within this approximation and neglecting the advective term in equation~(\ref{eq:f_rad}), we can write
\begin{equation}\label{eq:von_Zeipel_2}
    \boldsymbol{f}_{\rm rad} = \frac{L_{\rm rad}(\phi_{\rm R})}{L_{\rm Edd}(\phi_{\rm R})}  \boldsymbol{\nabla} \phi_{\rm R} \equiv \Gamma_{\rm Edd} (\phi_{\rm R}) \boldsymbol{\nabla} \phi_{\rm R},
\end{equation}
where $L_{\rm Edd} (\phi_{\rm R})$ is the donor's Eddington luminosity corresponding to a given $\phi_{\rm R}$ and we define the corresponding Eddington ratio $\Gamma_{\rm Edd} (\phi_{\rm R})$, i.e. the ratio of the radiative over Eddington luminosity at a given $\phi_{\rm R}$. 

The fundamental assumption underlying von Zeipel approximation is that the donor is barotropic, i.e. $\rho$, $T$, etc. are only functions of $\phi_{\rm R}$ \citep{vonZeipel1924} and that the effects of convection are ignored.  The application of this theorem to Roche-lobe-filling stars also seems problematic, because the von Zeipel theorem predicts zero radiative flux at the L1 point. We will further discuss the limitations in Section~\ref{sec:discussion}.

\section{Limiting cases}\label{sec:limiting}

In this section, we further simplify the 1D radiation hydrodynamic equations governing the MT through the nozzle around L1 that were introduced in Section~\ref{sec:1D}. The key assumption we use in all limiting cases is the von Zeipel theorem (Section~\ref{sec:von_zeipel}). We first show the hydrostatic case in Section~\ref{sec:hydrostatic}, which subsequently motivates the small-velocity (small-$v$) case ($v \ll c$) in Section~\ref{sec:small_v}. We then move to the adiabatic case where there is no energy exchange between gas and radiation in Section~\ref{sec:adiabatic}.

\subsection{Optically-thick hydrostatic case}\label{sec:hydrostatic}
In the limit of $v \to 0$ the co-moving frame becomes identical with the lab frame and the continuity equation~(\ref{eq:1D_mass}) now becomes trivial, $v = 0$. Also, $q_{\rm rad}$ (equation~\ref{eq:q_rad}) in equation~(\ref{eq:1D_en}) has to be finite, thus $E_{\rm rad} \to a_{\rm rad} T^4$. Assuming optically thick limit, i.e. $R \to 0$, $\lambda \to 1/3$, and $f \to 1/3$ (equation~\ref{eq:flux_lim}), we further obtain $P_{\rm rad} \to E_{\rm rad}/3$ in equation~(\ref{eq:3D_6}). Finally, using the von Zeipel theorem either in the form of equation~(\ref{eq:von_Zeipel}) or (\ref{eq:von_Zeipel_2}) and considering equations~(\ref{eq:1D_mom},\ref{eq:f_rad_1D}) leads to the system of six equations for six unknowns (three related to gas, $\rho$, $P_{\rm gas}$, $v$, and three to radiation $P_{\rm rad}$, $E_{\rm rad}$, $f_{\rm rad}$), where only two equations are non-trivial due to our approximations
\begin{subequations}\label{eq:12_0}
    \begin{align}
        & \frac{1}{\rho} \frac{\dd P_{\rm gas}}{\dd x} = - \lp 1 - \Gamma_{\rm Edd}\rp \frac{\dd \phi_{\rm R}}{\dd x}, \label{eq:12_0-1}  \\
        &  \frac{1}{\rho} \frac{\dd P_{\rm rad}}{\dd x} = - \Gamma_{\rm Edd} \frac{\dd \phi_{\rm R}}{\dd x}, \label{eq:12_0-2}
    \end{align}
\end{subequations}
with
\begin{equation}
    v = 0, \ P_{\rm rad} = \frac{1}{3} a_{\rm rad}T^4, \ E_{\rm rad} = a_{\rm rad}T^4, \ f_{\rm rad} = \Gamma_{\rm Edd} \frac{\dd \phi_{\rm R}}{\dd x}.
\end{equation}
Instead of equation~(\ref{eq:12_0-2}), we can equivalently use $F_{\rm rad} = {L}/({4 \pi G M}) ({\dd \phi_{\rm R}}/{\dd x})$. 

Using the above set of equations, we can in principle reconstruct part of the donor's hydrostatic envelope close to the L1 point, specifically, between some point $x_0$ and the L1 point located at $x_1$, $x_0 < x_1$. In order to close the system of equations we still need an EOS for $P_{\rm gas}$ and two boundary conditions: $\rho(x_0) = \rho_0$ and $T(x_0) = T_0$ that need to be taken from the underlying hydrostatic structure of a stellar evolution code. We also need to evaluate the Eddington ratio $\Gamma_{\rm Edd}$ (equation~\ref{eq:von_Zeipel_2}). The exact way to connect the $x$ coordinate with the potential $\phi$ and consecutively with the radius $r$ was explained in \citetalias{cehula2023} and will also be discussed later in Section~\ref{sec:relation_to_hydstat}.

\subsection{Optically-thick, $v \ll c$ case}\label{sec:small_v}
We can generalize the previous case to a situation where there is a slow but nonzero velocity, $v \ll c$. We apply assumptions similar to those in the hydrostatic case , but in order to obtain the continuity equation, we need a prescription for the cross section of the nozzle $Q$. In our previous work (\citetalias{cehula2023}, but see also \citealt{meyer1983}), we assumed $Q \propto c_{T, \rm gas}^2 \propto P_{\rm gas}/\rho$, which assumes that the gas is in adiabatic hydrostatic equilibrium in the plane perpendicular to to flow toward L1. For simplicity, we apply the same assumption here and we arrive at the following set of equations
\begin{subequations}\label{eq:1200}
    \begin{align}
        & \frac{1}{v} \frac{\dd v}{\dd x}  + \frac{1}{P_{\rm gas}} \frac{\dd P_{\rm gas}}{\dd x} = 0, \label{eq:1200_1} \\
        & v \frac{\dd v}{\dd x} + \frac{1}{\rho} \frac{\dd P_{\rm gas}}{\dd x} = - \lp 1 - \Gamma_{\rm Edd} \rp \frac{\dd \phi_{\rm R}}{\dd x}, \label{eq:1200_2} \\
        &  \frac{1}{\rho} \frac{\dd P_{\rm rad}}{\dd x} = -  \Gamma_{\rm Edd} \frac{\dd \phi_{\rm R}}{\dd x}, \label{eq:1200_3}
    \end{align}
\end{subequations}
with
\begin{equation}
    E_{\rm rad} = a_{\rm rad}T^4, \quad P_{\rm rad} = \frac{E_\text{rad}}{3}, \quad f_{\rm rad} = \frac{\kappa_{\rm R}}{c} F_{\rm rad} = \Gamma_{\rm Edd} \frac{\dd \phi_{\rm R}}{\dd x},
\end{equation}
where we have three non-trivial (differential) equations for three unknowns related to gas properties and the radiation unknowns are trivially (algebraically) related. This is a two-point boundary value problem. We have two boundary conditions at $x_0$, $\rho(x_0) = \rho_0$ and $T(x_0) = T_0$, which are fixed to values from a 1D hydrostatic model. The third boundary condition is applied at $x_{\rm crit}$, $v^2(x_{\rm crit}) = P_{\rm gas}(x_{\rm crit})/\rho(x_{\rm crit})$, which is the critical point either where $\Gamma_{\rm Edd} = 1$ or at the point L1, where $\dd \phi_{\rm R}/\dd x = 0$. The resulting MT rate is calculated using equation~(\ref{eq:Mdot}), where the exact value of $Q$ is calculated using an analytical estimate (equation~\ref{eq:Q}) derived later in Section~\ref{sec:analytic}.

\subsection{Optically-thick, adiabatic, $v \ll c$ case}\label{sec:adiabatic}
Let us now consider the same case as before, but instead of equation~(\ref{eq:1200_3}) we now assume that there is no energy exchange between radiation and gas, i.e. $q_{\rm rad} = 0$ in equation~(\ref{eq:1D_en}), and the gas flow is adiabatic. There can still be momentum exchange between gas and radiation. Thus, only the energy equation changes and we can write
\begin{subequations}\label{eq:121}
    \begin{align}
        & \frac{1}{v} \frac{\dd v}{\dd x}  + \frac{1}{P_{\rm gas}} \frac{\dd P_{\rm gas}}{\dd x} = 0, \label{eq:121_1} \\
        & v \frac{\dd v}{\dd x} + \frac{1}{\rho} \frac{\dd P_{\rm gas}}{\dd x} = - \lp 1 -  \Gamma_{\rm Edd}\rp \frac{\dd \phi_{\rm R}}{\dd x}, \label{eq:121_2} \\
        & \frac{\dd \epsilon_{\rm gas}}{\dd x} - \frac{P_{\rm gas}}{\rho} \frac{1}{\rho} \frac{\dd \rho}{\dd x} = 0, \label{eq:121_3}
     \end{align}  
\end{subequations}
with
\begin{equation}
    E_{\rm rad} = a_{\rm rad}T^4, \quad P_{\rm rad} = \frac{E_\text{rad}}{3}, \quad f_{\rm rad} = \frac{\kappa_{\rm R}}{c} F_{\rm rad} = \Gamma_{\rm Edd} \frac{\dd \phi_{\rm R}}{\dd x},
\end{equation}
and the same boundary conditions as in the previous case (Sec.~\ref{sec:small_v}, equation~\ref{eq:1200}).

\section{Sub-Eddington mass transfer}\label{sec:subedd}

Our goal in this Section is to provide solutions to equations outlined in Section~\ref{sec:theory} and \ref{sec:limiting} under the assumption that the MT is sub-Eddington. Specifically, this means that when we follow the solution from an interior point, the first critical point we encounter is due to the potential flattening at L1. We discuss situations where $\Gamma_\text{Edd} >1$ inside of L1 in Section~\ref{sec:superedd}.

This Section begins by first discussing how to match our solutions to an underlying hydrostatic stellar model (Section~\ref{sec:relation_to_hydstat}). Then we proceed to formulate analytic estimates of the relevant quantities to develop an intuitive understanding of the expected results (Section~\ref{sec:analytic}). After that, we present several cases where the equations permit algebraic solutions, which are possible for isothermal and ideal gas assumptions relevant for optically thin and optically thick situations (Section~\ref{sec:algebraic}). Next, we present numerical solutions with a realistic EOS and illustrate the results on two massive stars undergoing MT (Section~\ref{sec:real_EOS}). Finally, we summarize our results and compare them with previous work (Section~\ref{sec:comparison}).

\subsection{Relation to hydrostatic structure} \label{sec:relation_to_hydstat}
An important aspect of our model is the connection between the hydrodynamical mass flow around L1 and the underlying hydrostatic donor envelope structure calculated using a stellar evolution code. Similarly to \citetalias{cehula2023}, we notice that all the limiting cases of our model that we are able to solve, sets of equations~(\ref{eq:1200}) and (\ref{eq:121}), can be reformulated so that the Roche potential $\phi_{\rm R}$ becomes the independent variable instead of the coordinate $x$, allowing us to completely omit the coordinate $x$. This is what we do in practice in this paper. The mapping between $\phi$ and $r$ is based on the assumption that equipotentials can be related to spheres with the same equivalent volume. In this work, we use the higher order approximation derived by \citet{jackson2017} to make the correspondence, see equation~(\ref{eq:phi_vol}). We denote the radius corresponding to the inner boundary of our model at $\phi_0$, or also $x_0$, by $R_0$. The radius corresponding to the equipotential passing through the L1 point with $\phi_1$ is $R_\text{L}$.

If the photosphere is above the L1 point, $R_{\rm d} > R_{\rm L}$, we have a Roche-lobe-overflowing donor and a regime that has traditionally been labeled as optically thick. In this case, we calculate the potential difference corresponding to $R_0$ as
\begin{equation}\label{eq:phi_thick}
    \phi_1 - \phi_0 \equiv \int_{R_0}^{R_{\rm{L}}} \dd \bar{\phi} = - \int_{R_0}^{R_{\rm{L}}} \frac{\dd \bar{P}}{\bar{\rho}},
\end{equation}
where $\bar{P}$ is the hydrostatic total pressure inside of the donor. 

If the photosphere is below the L1 point, $R_{\rm d} < R_{\rm L}$, we have a Roche-lobe-underfilling donor and an optically thin regime. In this case, we take the inner boundary of our model to be the donor's photosphere (unlike \citetalias{cehula2023}) and we calculate the potential difference corresponding to $R_0 = R_{\rm d}$ according to
\begin{equation}\label{eq:phi_thin}
    \phi_1 - \phi_0 \equiv \phi_V (R_{\rm{L}}) - \phi_V( R_{\rm d}),
\end{equation}
where $\phi_V$ is the equipotential reached by a donor of radius $r_V$.

In the optically thick regime, it is useful to express the distance between $R_0$ and $R_{\rm{L}}$ (or equivalently between $\phi_0$ and $\phi_1$, or $x_0$ and $x_1$) by the number of pressure scale heights $\Delta N_{H_P}$. We define the pressure scale height number $N_{H_P}$ for the donor's interior by
\begin{equation}
    N_{H_P} (r) \equiv \int_{r}^{R_{\rm{d}}} \frac{\dd r^\prime}{H_P (r^\prime)}, \quad \text{for} \quad  r \leq R_\dd
\end{equation}
where $H_P$ is the pressure scale height. We see that $N_{H_P} (R_{\rm d}) = 0$ and that $N_{H_P}$ increases inward and decreases outward. Hence, for a Roche-lobe-overflowing donor we have
\begin{equation}
    \Delta N_{H_P} = N_{H_P} (R_0) - N_{H_P} (R_{\rm{L}}) = \int_{R_0}^{R_{\rm{L}}} \frac{\dd r}{H_P}.
    \label{eq:DN_HP_thick}
\end{equation}

Finally, let us also explain how we calculate the Eddington factor $\Gamma_{\rm Edd}$. The usage of the von Zeipel theorem does not change the value of the Eddington factor. Thus, in the optically thick regime in the donor's interior we calculate it according to \citep[e.g.][]{sanyal2015}
\begin{equation}\label{eq:Gamma_Edd}
    \Gamma_{\rm Edd} = \frac{L_{\rm rad}}{L_{\rm Edd}} = \frac{4 \bar{P}_{\rm rad}}{\bar{P}} \bar{\nabla}_T,
\end{equation}
where $\bar{P}_{\rm rad}$ is the hydrostatic radiative pressure and $\bar{\nabla}_T$ is the actual hydrostatic temperature gradient. In the optically thin regime outside the donor we use a simple model, where we assume $\Gamma_{\rm Edd}$ to be constant and take the value at the photosphere.

\subsection{Analytic estimates}\label{sec:analytic}
The sets of equations~(\ref{eq:1200}) or (\ref{eq:121}) cannot be analytically solved for a general EOS, but we can still analytically estimate the quantity of main interest, $-\dot{M}_{\rm d} = Q\rho v$ (equation~\ref{eq:Mdot}). We need to know the underlying hydrostatic structure of the donor from a 1D stellar evolution code, such as the hydrostatic stellar quantities $\bar{\rho}$, $\bar{P}_{\rm gas}$ as a function of stellar radius $r$ or alternatively the stellar gravitational potential $\bar{\phi}$. We also need to know the gas cross-section $Q$. We need to evaluate $\dot{M}_{\rm d}$ at one point, and the easiest choice is the critical point located where the right-hand side of equation~(\ref{eq:1200_2}) vanishes. If the Eddington limit is not reached, then the critical point occurs at the Roche-lobe radius $R_{\rm L}$ that corresponds to $\bar{\phi} (R_{\rm L}) = \phi_1$, where $\phi_1$ is the Roche potential at L1. Combining equations~(\ref{eq:1200_1}) and (\ref{eq:1200_2}) we see that the critical velocity $v_{\rm crit} \equiv v(x_{\rm crit})$ is given by $v_{\rm crit}^2 = c_{T, \rm{gas}}^2 = P_{\rm gas}/\rho$. Now, we need to express the hydrodynamic values $P_{\rm gas}$, $\rho$, which we a priori do not know, as functions of the hydrostatic values $\bar{P}_{\rm gas}$, $\bar{\rho}$, which we know from a stellar evolution code. Considering the simple case of an isothermal gas (see Sec. 5.1 in \citetalias{cehula2023}, or also Sec. 3.1.3 in \citealt{lamers1999}) we can approximately write $P_{\rm gas} = \bar{P}_\text{gas}/\sqrt{\text{e}}$ and $\rho = \bar{\rho} / \sqrt{\rm e}$ at the critical point. 

In order to approximate $Q$ we consider the lowest order approximation of the Roche potential in the $yz$-plane
 \begin{equation}
    \phi_{\rm R}(x, y, z) = \phi_{\rm R}^x(x) + \frac{B}{2} y^2 + \frac{C}{2} z^2,
\end{equation}
where the centers of the two stars lie on the $x$ axis and the orbital plane is the $xy$ plane \citepalias{cehula2023}. The cross-section $Q$ can then be approximated as
\begin{equation}\label{eq:Q}
    Q \equiv \left. \frac{\dd Q}{\dd \phi}\right\vert_{\rm{L1}} \frac{P_{\rm gas}}{\rho}.
\end{equation}
If we assume $\Gamma_{\rm Edd}$ to be roughly constant in the $yz$-plane then the potential is effectively lower by a factor of $(1-\Gamma_{\rm Edd})$ and it is equivalent to assume $B \to (1-\Gamma_{\rm Edd})B$ and $C \to (1-\Gamma_{\rm Edd})C$. In such a case we can write (see the derivation of $Q_\rho$ in \citetalias{cehula2023})
\begin{equation}\label{eq:dQ_dphi}
    \left. \frac{\dd Q}{\dd \phi}\right\vert_{\rm{L1}} = \left. \frac{1}{1 - \Gamma_{\rm Edd}} \frac{2 \pi}{\sqrt{BC}} \right|_{r = R_{\rm L}}.
\end{equation}
We see that $Q$ diverges for $\Gamma_{\rm Edd} \to 1$.

Nevertheless, if the Eddington limit is not reached and using the definition of $Q$ in equations~(\ref{eq:Q}) and (\ref{eq:dQ_dphi}), $\dot{M}_{\rm d}$ in equation~(\ref{eq:Mdot}) can be approximated as
\begin{equation}\label{eq:dotM_an,thick}
    -\dot{M}_{\rm d} \approx \dot{M}_{\rm an, thick} \equiv \left. \frac{1}{1 - \Gamma_{\rm Edd}} \frac{1}{\sqrt{e}} \frac{2 \pi}{\sqrt{BC}} \frac{\bar{P}_{\rm gas}^{3/2}}{\bar{\rho}^{1/2}} \right|_{r = R_{\rm L}}.
\end{equation}
This expression can be evaluated only for $r \leq R_{\rm d}$ and therefore applies only to a Roche-lobe-overflowing donor ($R_{\rm L} \leq R_{\rm d}$). Consequently, we follow tradition and refer to this quantity as the optically thick MT rate. We will derive a separate expression for $r \geq R_{\rm d}$ in Section~\ref{sec:iso}, equation~(\ref{eq:dotM_an,thin}).

\subsection{Algebraic solutions}\label{sec:algebraic}
We obtain algebraic solutions for two specific simplifications: isothermal gas (Section~\ref{sec:iso}), which provides a solution in the optically thin regime,  and ideal gas (Section~\ref{sec:id}).

\subsubsection{Isothermal gas}\label{sec:iso}
Let us first consider a simple case where $P_{\rm gas}$ is given by ideal gas law (equation~\ref{eq:id}) and the temperature $T$ is constant. In this case, the nozzle cross-section $Q$ given by equation~(\ref{eq:Q}) is constant and the sets of equations~(\ref{eq:1200}) or (\ref{eq:121}) reduce to
\begin{subequations}\label{eq:iso}
    \begin{align}
	& \frac{\dd \ln v}{\dd x} + \frac{\dd \ln \rho}{\dd x} = 0, \label{eq:iso_mass} \\
	& v^2\frac{\dd \ln v}{\dd x} + c_{T, \rm gas}^2 \frac{\dd \ln \rho}{\dd x} = -(1 - \Gamma_{\rm Edd}) \frac{\dd\phi_{\rm R}}{\dd x}, \label{eq:iso_mom} \\
	& T = T_0 = \rm{const}, \label{eq:iso_en}
    \end{align}
\end{subequations}
where instead of the energy equation~(\ref{eq:1200_3}) or (\ref{eq:121_3}) we assume $T$ is constant. The algebraic solution of this set of equations can be written in the form
\begin{subequations}
    \begin{align}
	& \frac{1}{2} \left[ \lp \frac{v(x)}{c_{T, \rm gas}}\rp^2 - \lp \frac{v_0}{c_{T, \rm gas}}\rp^2 \right] - \ln \frac{v(x)}{v_0} = - \lp 1 - \Gamma_{\rm Edd}\rp \frac{\phi_{\rm R}^x (x) - \phi_0}{c_{T, \rm gas}^2}, \\
	& \frac{\rho(x)}{\rho_0} = \frac{v_0}{v(x)},
    \end{align}
\end{subequations}
where we start at $x = x_0$, $\phi_0 = \phi_{\rm R}^x(x_0)$, $\rho_0 = \rho(x_0)$, $v_0 \equiv v (x_0)$, and we end at $x = x_1$ (the L1 point), provided that $\Gamma_{\rm Edd} < 1$, where $v(x_1) = c_{T, \rm gas}$. The velocity $v_0$ is the solution of the algebraic equation
\begin{equation}
    \frac{1}{2} \lp \frac{v_0}{c_{T, \rm gas}} \rp^2 - \ln \frac{v_0}{c_{T, \rm gas}} = \lp 1 - \Gamma_{\rm Edd}\rp \frac{\phi_1-\phi_0}{c_{T, \rm gas}^2} + \frac{1}{2}.
\end{equation}

The only difference between this isothermal limit and the isothermal limit in \citetalias{cehula2023} is that the solution here includes the momentum exchange between gas and radiation. Therefore, the potential difference the gas has to cover between $x_0$ and $x_1$ (L1) is multiplied by a factor of $(1- \Gamma_{\rm Edd})$ and thus somewhat reduced: $(\phi_1-\phi_0)/c_{T, \rm gas}^2 \to (1-\Gamma_{\rm Edd})(\phi_1-\phi_0)/c_{T, \rm gas}^2$. The isothermal case in \citetalias{cehula2023} is equivalent to the optically thin MT rate of \citet{ritter1988}, or its improved version of \citet{jackson2017}, depending on how we choose to approximate the difference between the potentials at the L1 point $\phi_1$ and the photosphere $\phi_{\rm ph}$. Thus, the MT rate corresponding to the isothermal case with radiation is
\begin{equation}\label{eq:dotM_an,thin}
    \begin{aligned}
        - \dot{M}_{\rm d} &= \dot{M}_{\rm an, thin} \equiv \dot{M}_0 \exp \left[ - \lp 1 - \Gamma_{\rm Edd}\rp \frac{\phi_1 - \phi_{\rm ph}}{c_{T, \rm gas}^2}\right], \\
        & \text{where} \quad \dot{M}_0 = \frac{1}{1-\Gamma_{\rm Edd}} \frac{1}{\sqrt{\rm e}} \frac{2 \pi}{\sqrt{BC}} \frac{P_{\rm gas, ph}^{3/2}}{\rho_{\rm ph}^{1/2}},
    \end{aligned}
\end{equation}
where $P_{\rm gas,ph}$ and $\rho_{\rm ph}$ are the photosphere gas pressure and density, respectively. We will use this optically thin analytical estimate for Roche-lobe-underfilling donors, $R_{\rm d} \leq R_{\rm L}$ and  $\phi_{\rm ph} \leq \phi_1$, for example in Section~\ref{sec:40Msun}.

It should be noted that the factor $1-\Gamma_\text{Edd}$ is present not only in the expression for $\dot{M}_0$, but also in the exponential factor in equation~(\ref{eq:dotM_an,thin}), leading to the exponential dependence of $\dot{M}_\text{d}$ on $1-\Gamma_\text{Edd}$. For example, for a photosphere located about four scale heights below L1, $(\phi_1 - \phi_\text{ph})/c_{T,\text{gas}}^2 \approx 4$, an Eddington factor of $\Gamma_\text{Edd}=0.5$ will lead to an increase in $\dot{M}_\text{d}$ by a factor of $2 e^2 \approx 15$. This means that Roche-lobe-underfilling stars with high $\Gamma_\text{Edd}$ will exhibit significantly increased $\dot{M}_\text{d}$  compared to existing prescriptions.

\subsubsection{Ideal gas}\label{sec:id}
Here, we assume an ideal gas EOS (equation~\ref{eq:id}) for $P_{\rm gas}$. However, applying this assumption in the small-$v$ case (Section~\ref{sec:small_v}, equations~\ref{eq:1200}), does not yield an algebraic solution. Instead, we focus on the adiabatic case (Section~\ref{sec:adiabatic}, equations~\ref{eq:121}), which gives
\begin{subequations}\label{eq:hydro_id}
	\begin{align}
	   & \frac{\dd \ln v}{\dd x} + \frac{\dd \ln \rho}{\dd x} + \frac{\dd \ln T}{\dd x} = 0, \label{eq:hydro_id_mass} \\
	   & v^2\frac{\dd \ln v}{\dd x} + \gamma c_{T, \rm gas}^2 \frac{\dd \ln \rho}{\dd x} = - (1- \Gamma_{\rm Edd})\frac{\dd\phi_{\rm R}}{\dd x}, \label{eq:hydro_id_mom} \\
	   & \frac{\dd \ln \rho}{\dd x} - \frac{1}{\gamma-1} \frac{\dd \ln T}{\dd x} = 0. \label{eq:hydro_id_en}
	\end{align}
\end{subequations}
Again as before, the inner boundary is at $x = x_0$ and the outer at $x = x_1$, i.e. the L1 point, provided that $\Gamma_{\rm Edd} < 1$. The boundary conditions are $\rho(x_0) = \rho_0$, $T(x_0) = T_0$, and $v^2(x_1) = c_{T, \rm gas}^2 (x_1) = kT(x_1)/\overline{m}$. The above set of equations can be solved algebraically,
\small
\begin{subequations}
    \begin{align}
	    &\frac{1}{2} \left[ \lp \frac{v}{c_0}\rp^2 - \lp\frac{v_0}{c_0}\rp^2 \right] + \frac{\gamma}{\gamma-1} \left[\lp \frac{v_0}{v}\rp^{\frac{\gamma-1}{\gamma}} - 1\right] = - \lp 1- \Gamma_{\rm Edd}\rp \frac{\phi_{\rm R}^x - \phi_0}{c_0^2}, \\
	    & \rho = \lp \frac{v_0}{v }\rp^{\frac{1}{\gamma}} \rho_0, \\
        & T = \lp\frac{v_0}{v}\rp^\frac{\gamma-1}{\gamma} T_0,
    \end{align}
\end{subequations}
\normalsize
where $c_0^2 \equiv c_{T, \rm gas}^2 (x_0) = k T_0 / \overline{m}$ and $v_0$ is the solution of
\small
\begin{equation}
    \frac{3\gamma-1}{2(\gamma-1)} \lp \frac{v_0}{c_0}\rp^{\frac{2(\gamma-1)}{3\gamma-1}} - \frac{1}{2} \lp \frac{v_0}{c_0}\rp^2 = - \lp 1 - \Gamma_{\rm Edd}\rp\frac{\phi_1-\phi_0}{c_0^2} + \frac{\gamma}{\gamma-1}.
\end{equation}
\normalsize
As before, compared to the ideal gas limit of \citetalias{cehula2023}, the only difference is that the current ideal gas case includes momentum exchange between gas and radiation. Consequently, the potential difference is reduced: $(\phi_1-\phi_0)/c_{T, \rm gas}^2 \to (1-\Gamma_{\rm Edd})(\phi_1-\phi_0)/c_{T, \rm gas}^2$. Therefore, we do not discuss this case further in this paper and refer the reader to \citetalias{cehula2023} for more details.

\subsection{Numerical solution for a realistic equation of state}\label{sec:real_EOS}
For a general EOS, we solve the set of hydrodynamic equations~(\ref{eq:1200}) or (\ref{eq:121}) using the same framework we developed in \citetalias{cehula2023} based on a relaxation method of \citet[Chapter 18]{press2007}. The realistic EOS is adopted from the MESA \texttt{eos} module \citep{paxton2011, paxton2013, paxton2015, paxton2018, paxton2019}. Depending on the metallicity $Z$ and the position in the $\rho T$ plane, one of the following components is used: OPAL/SCVH \citep{rogers2002,saumon1995}, Free EOS \citep{irwin2008},  HELM \citep{timmes2000}, PC \citep{potekhin2010}, Skye \citep{jermyn2021}, or CMS \citep{chabrier2019}. Following the tutorial by \citet{timmes2021}, we construct a table with all relevant thermodynamic variables. We interpolate in the table using bilinear interpolation \citep[Chapter 3]{press2007}. We refer the reader to our original work \citepalias{cehula2023} for more details. 

In what follows, we compare numerical solutions in the small-$v$ case (Section~\ref{sec:small_v}, equations~\ref{eq:1200}), the adiabatic case (Section~\ref{sec:adiabatic}, equations~\ref{eq:121}) and our original model in \citetalias{cehula2023}. First, in Section~\ref{sec:30Msun} we use the same low metallicity donor $30\,\rm{M}_\odot$ as in \citetalias{cehula2023} that is adopted from \citet{marchant2021}. Our results will justify the usage of the analytical estimate in the optically thick regime (equation~\ref{eq:dotM_an,thick}). Next, in Section~\ref{sec:40Msun} we illustrate the differences between our present approach and the standard approach on a $40\,\rm{M}_\odot$ low-metallicity donor in a supergiant stage. This donor is ideal for this purpose because it achieves high values of the Eddington ratio in the surface layers ($\Gamma_{\rm Edd} > 0.5$), but does not cross into the super-Eddington regime. For completeness, we present in Appendix~\ref{app:1Msun} comparison for a $1\,\rm{M}_\odot$ red giant showing that in this case radiation has little effect on $\dot{M}$, as expected.

Finally, we introduce the notation for the previous MT rate prescriptions. For Roche-lobe-underfilling donors, $R_{\rm d} \leq R_{\rm L}$, we will use the optically thin model of \citet{jackson2017}, which is based on \citet{ritter1988}. We will refer to this MT rate by $\dot{M}_{\rm J}$ and the expression can be found in equation~(\ref{eq:Mdot_J}). For a Roche-lobe-overflowing donor, $R_{\rm d} \geq R_{\rm L}$, we will use the optically thick model of \citet{kolb1990}. We will refer to this MT rate by $\dot{M}_{\rm KR}$ and the expression is in equation~(\ref{eq:Mdot_KR}). In some cases, we will also use the prescription introduced by \citet{marchant2021}, $\dot{M}_{\rm M}$.

\subsubsection{30 $\rm{M}_\odot$ donor undergoing thermal time-scale mass transfer}\label{sec:30Msun}
We used a $30\,\rm{M}_\odot$ low-metallicity donor undergoing intensive thermal time-scale MT to a black hole, as was investigated by \citet{marchant2021}. The donor's initial metallicity is $Z = \rm{Z}_\odot/10$, where $\rm{Z}_\odot = 0.0142$ \citep{asplund2009} and where the relative metal mass fractions are from \citet{grevesse1998}. We take a model with a $7.5\,\rm{M}_\odot$ black hole and an initial period of $31.6$~d. We recalculated the evolution using the full MT prescription $\dot{M}_{\rm{M}}$ and the same MESA version as in \citet{marchant2021}.

\begin{figure}
    \centering
    \includegraphics[width=0.99\columnwidth]{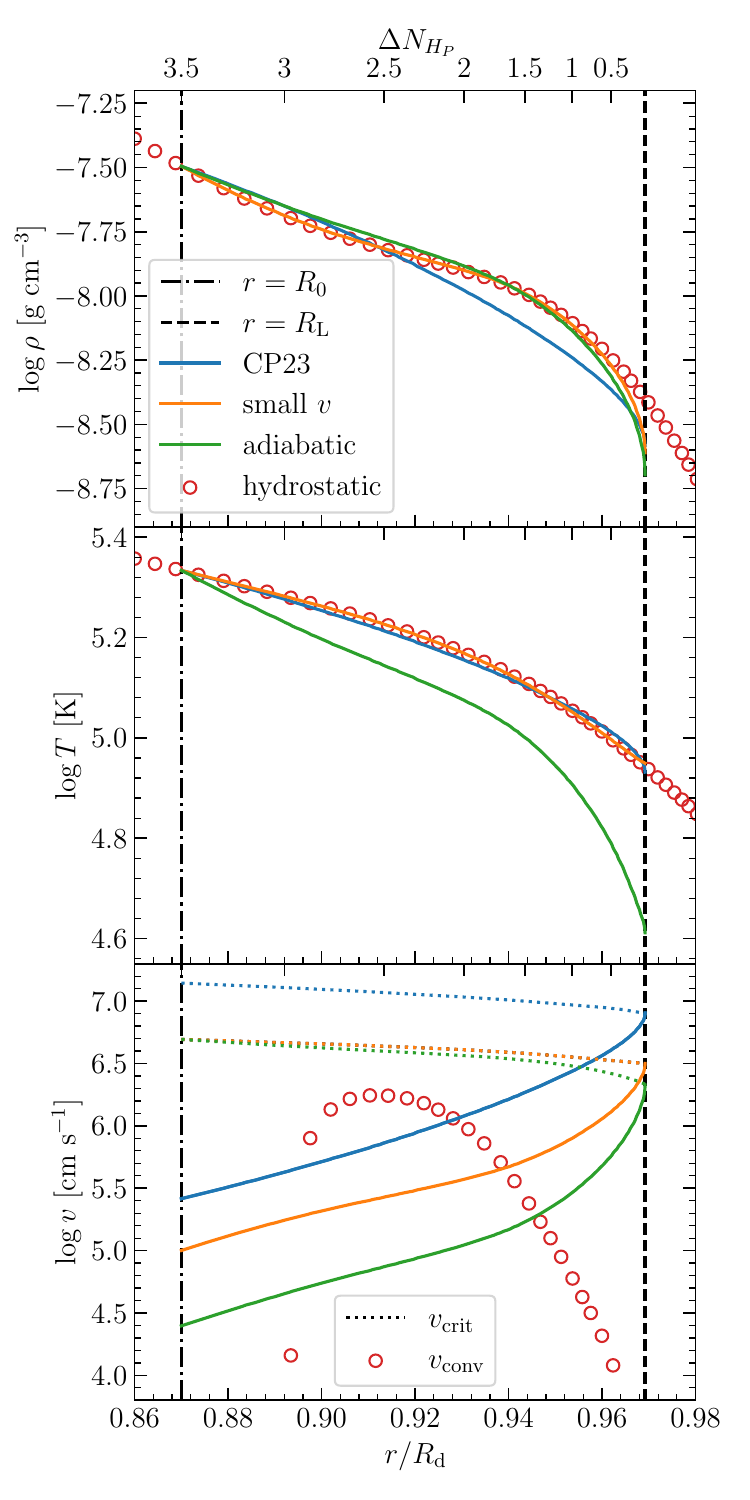}
    \caption{Comparison of density $\rho$, temperature $T$, and velocity $v$ profiles for a $30 \rm{M}_\odot$ low-metallicity star undergoing thermal time-scale MT \citepalias{cehula2023}, small-$v$ case (Section~\ref{sec:small_v}), and adiabatic case (Section~\ref{sec:adiabatic}). We also show the underlying MESA hydrostatic density and temperature profiles and the value of convective velocity $v_{\rm conv}$. The profiles are shown as functions of radius $r$ going from the inner boundary at $R_0$ (that corresponds to $x_0$), to the outer boundary at  $R_{\rm L}$ (the L1 point at $x_1$). The critical velocity $v_{\rm crit}$, indicated by dotted lines, is reached at the critical point, L1. With the top $x$ axis we indicate the distance to the critical point in units of pressure scale heights $\Delta N_{H_P}$.}
    \label{fig:prof-comp-m7}
\end{figure}

As in \citetalias{cehula2023}, we first examine the binary near the end of the thermal time-scale MT phase at an age of $6.79\,\rm{Myr}$, with a surface Eddington factor $\Gamma_{\rm Edd} = 0.78$, donor mass $M_{\rm{d}} = 14.0\,\rm{M}_\odot$, core helium abundance $0.36$, relative radius excess $\delta R_\text{d} = 3.18 \times10^{-2}$, and mass ratio $q = 1.85$. Here, $\delta R_{\rm d} \equiv \Delta R_{\rm d}/R_{\rm L}$, where $\Delta R_{\rm d} \equiv R_{\rm d} - R_{\rm L}$ is the radius excess, and $q \equiv M_{\rm d} / M_{\rm a}$ is the donor-to-accretor (black hole) mass ratio.  

In Fig.~\ref{fig:prof-comp-m7}, we compare the density, temperature, and velocity profiles for a fixed value of $R_0$ with the hydrostatic values from MESA. We find that the small-$v$ case best matches the MESA hydrostatic profile, reproducing it well up to the vicinity of the L1 point, where the hydrostatic assumption $v \ll c_{T, \rm gas}$ in MESA breaks down. The only potential source of discrepancy is convection, which is not included in our model. However, even though convective velocities $v_{\rm conv}$ exceed flow velocities locally, our model still captures the MESA profile accurately.  

\begin{figure}
    \centering
    \includegraphics[width=0.99\columnwidth]{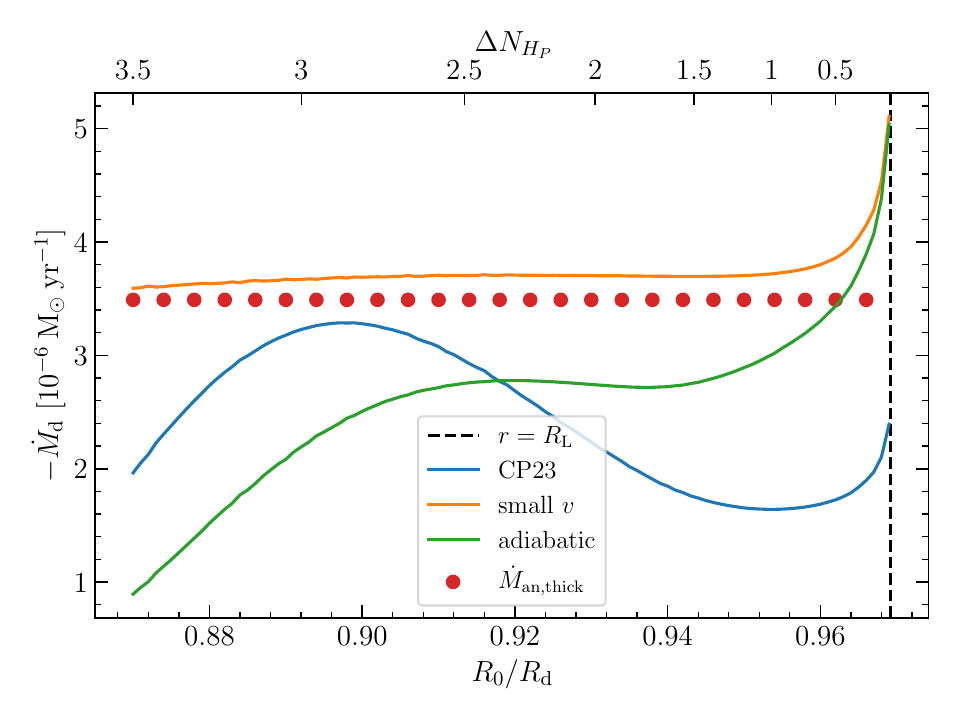}
    \caption{Comparison of the dependencies of the donor's MT rate $-\dot{M}_{\rm d}$ on $R_0$ for the same cases and donor star as in Fig.~\ref{fig:prof-comp-m7}. We also show the analytical estimate $\dot{M}_{\rm an,thick}$ (equation~\ref{eq:dotM_an,thick}). The outer fixed boundary at $R_L$ is indicated as well as the distance between $R_0$ and $R_{\rm{L}}$ in units of pressure scale heights $\Delta N_{H_P}$.}
    \label{fig:Mdot-R0-comp-m7}
\end{figure}

This agreement is also evident in Fig.~\ref{fig:Mdot-R0-comp-m7}, which shows the corresponding MT rates for different values of $R_0$. Furthermore, we include the analytical estimate $\dot{M}_{\rm an,thick}$ (equation~\ref{eq:dotM_an,thick}). We find that $\dot{M}_{\text{small},v}$ remains nearly constant up to the immediate vicinity of the L1 point $(\Delta N_{H_P} \lesssim 0.5)$, however, it is not surprising that the model becomes inaccurate when $R_0$ is positioned too close to the critical point. The small deviations arise from convective effects, which are not included in our model.  

In Fig.~\ref{fig:Mdot-R0-comp-m7} we also see that the MT rates of \citetalias{cehula2023} and in the adiabatic case follow a similar trend, but the \citetalias{cehula2023} model gives a lower value of the MT rate, which we discuss in more detail in Section~\ref{sec:comparison}. The similarity of the trends is caused by the fact that the \citetalias{cehula2023} model is also adiabatic (in the sense that there is no energy exchange between gas and radiation), but it also includes the averaging of physical quantities over the L1 plane. Therefore, as the gas approaches L1, it cools down and its cross-section decreases, its potential energy decreases, and the difference is used to heat up the gas, as we can also see in the middle panel of Fig.~\ref{fig:prof-comp-m7}.

\begin{figure}
    \centering
    \includegraphics[width=0.95\columnwidth]{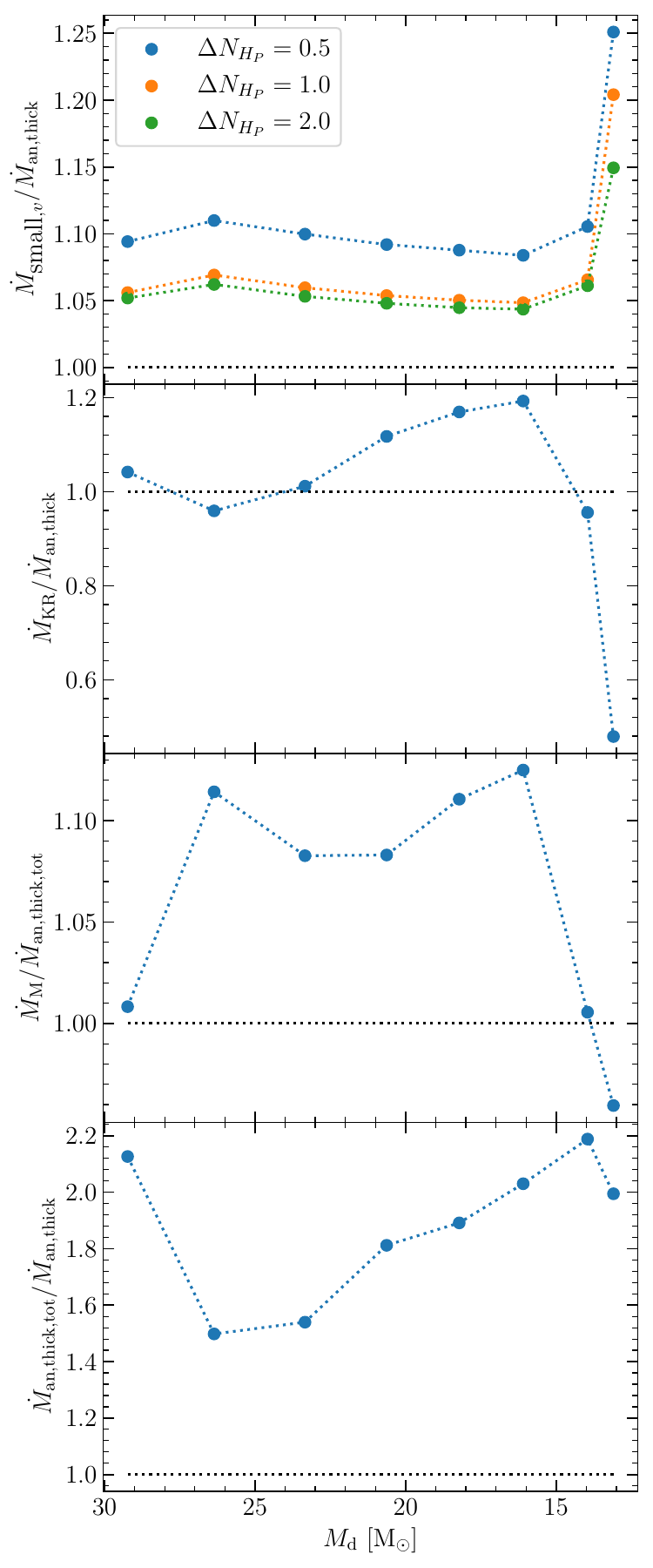}
    \caption{Comparison of different MT prescriptions throughout the evolution of originally $30 \rm{M}_\odot$ low-metallicity donor undergoing thermal time-scale MT (Section~\ref{sec:30Msun}, \citealt{marchant2021}). First panel: the small-$v$ case (Section~\ref{sec:small_v}) for three different values of the distances between the inner and outer boundary in units of pressure scale heights $\Delta N_{H_P}$ vs. the analytical estimate $\dot{M}_{\rm an,thick}$ (equation~\ref{eq:dotM_an,thick}). Second panel: the MT prescription introduced by \citet{kolb1990}, $\dot{M}_{\rm KR}$ vs. the analytical estimate $\dot{M}_{\rm an,thick}$. Third panel: the MT prescription introduced by \citet{marchant2021}, $\dot{M}_{\rm M}$, vs. the analytical estimate $\dot{M}_{\rm an,thick,tot}$ (equation~\ref{eq:dotM_an,thick,tot}). Fourth panel: Comparison of the two analytical prescriptions $\dot{M}_{\rm an,thick,tot}$ vs. $\dot{M}_{\rm an,thick}$, throughout the evolution.}
    \label{fig:Mdot-comp}
\end{figure}

In Fig.~\ref{fig:Mdot-comp}, we compare different MT rates throughout the evolution of the $30\,\rm{M}_\odot$ donor. First, we compare the small-$v$ case with the analytical expression $\dot{M}_{\rm an,thick}$ (equation~\ref{eq:dotM_an,thick}). We then compare the MT prescription of \citet{kolb1990}, $\dot{M}_{\rm KR}$, with $\dot{M}_{\rm an,thick}$, followed by a comparison of the MT rate $\dot{M}_{\rm M}$ from \citet{marchant2021} with the modified analytical expression  
\begin{equation}\label{eq:dotM_an,thick,tot}
    -\dot{M}_{\rm d} \approx \dot{M}_{\rm an, thick,tot} \equiv \left. \frac{1}{\sqrt{e}} \frac{2 \pi}{\sqrt{BC}} \frac{\bar{P}^{3/2}}{\bar{\rho}^{1/2}} \right|_{r = R_{\rm L}},
\end{equation}  
which modifies $\dot{M}_{\rm an,thick}$ by replacing $P_{\rm gas}$ with $P$ and removing the super-Eddington boost. Finally, we compare $\dot{M}_{\rm an,thick,tot}$ with $\dot{M}_{\rm an,thick}$.  

We find that the small-$v$ case and $\dot{M}_{\rm an,thick}$ agree well throughout the evolution, with differences within $\sim 20\%$. When the inner boundary is significantly below the L1 point ($\Delta N_{H_P} \lesssim 1$), $\dot{M}_{\text{small},v}$ remains nearly independent of $R_0$, the only free parameter in our model. The only exception occurs at the final point, where a small discrepancy arises between the orange and green markers, likely due to convection, which is not included in our model. Here, $v_{\rm conv}$ increases to $\sim 10^7\,\rm{cm}\,\rm{s}^{-1}$ below the surface. This agreement supports the use of $\dot{M}_{\rm an,thick}$ (equation~\ref{eq:dotM_an,thick}) as a benchmark for other MT prescriptions.  

In the second panel, $\dot{M}_{\rm an,thick}$ agrees with $\dot{M}_{\rm KR}$ to within $\sim$20\%, except at the final point, where $\dot{M}_{\rm an,thick}$ is approximately twice as large. In the third panel, $\dot{M}_{\rm an,thick,tot}$ matches $\dot{M}_{\rm M}$ within $\sim$10\%, further justifying its use. Finally, the bottom panel shows that $\dot{M}_{\rm an,thick}$ is consistently about twice as low as $\dot{M}_{\rm an,thick,tot}$, which would impact binary evolution if used self-consistently instead of $\dot{M}_{\rm an,thick,tot}$ ($\dot{M}_{\rm M}$). The effects would resemble those of evolving the binary with $\dot{M}_{\rm KR}$ (see the second panel of Fig.~\ref{fig:Mdot-comp}). For details, refer to \citet{marchant2021}, or see \citetalias{cehula2023}, where we previously illustrated the impact of reducing the MT rate by a factor of a few. Therefore, we do not repeat this analysis here but discuss the implications in more detail in Section~\ref{sec:discussion}.  

\subsubsection{40 $\rm{M}_\odot$ supergiant donor}\label{sec:40Msun}
Here, we use a single nonrotating star with an initial mass of $40\,\rm{M}_\odot$ and the same low metallicity as in Section~\ref{sec:30Msun}. We evolve the star to an age of 5.30 Myr, when we obtain a supergiant with a core helium fraction of 0.64, mass $M_{\rm d} = 38.2 \: \rm{M}_\odot$, effective temperature $T_{\rm eff} = 1.2 \times 10^4\,\rm{K}$, radius $R_{\rm d} = 190\,\rm{R}_\odot$, luminosity $L_{\rm d} = 6.5 \times 10^5\,\rm{L}_\odot$, and Eddington ratio $\Gamma_{\rm Edd} = 0.70$ at the surface.

\begin{figure}
    \centering
    \includegraphics[width=0.99\columnwidth]{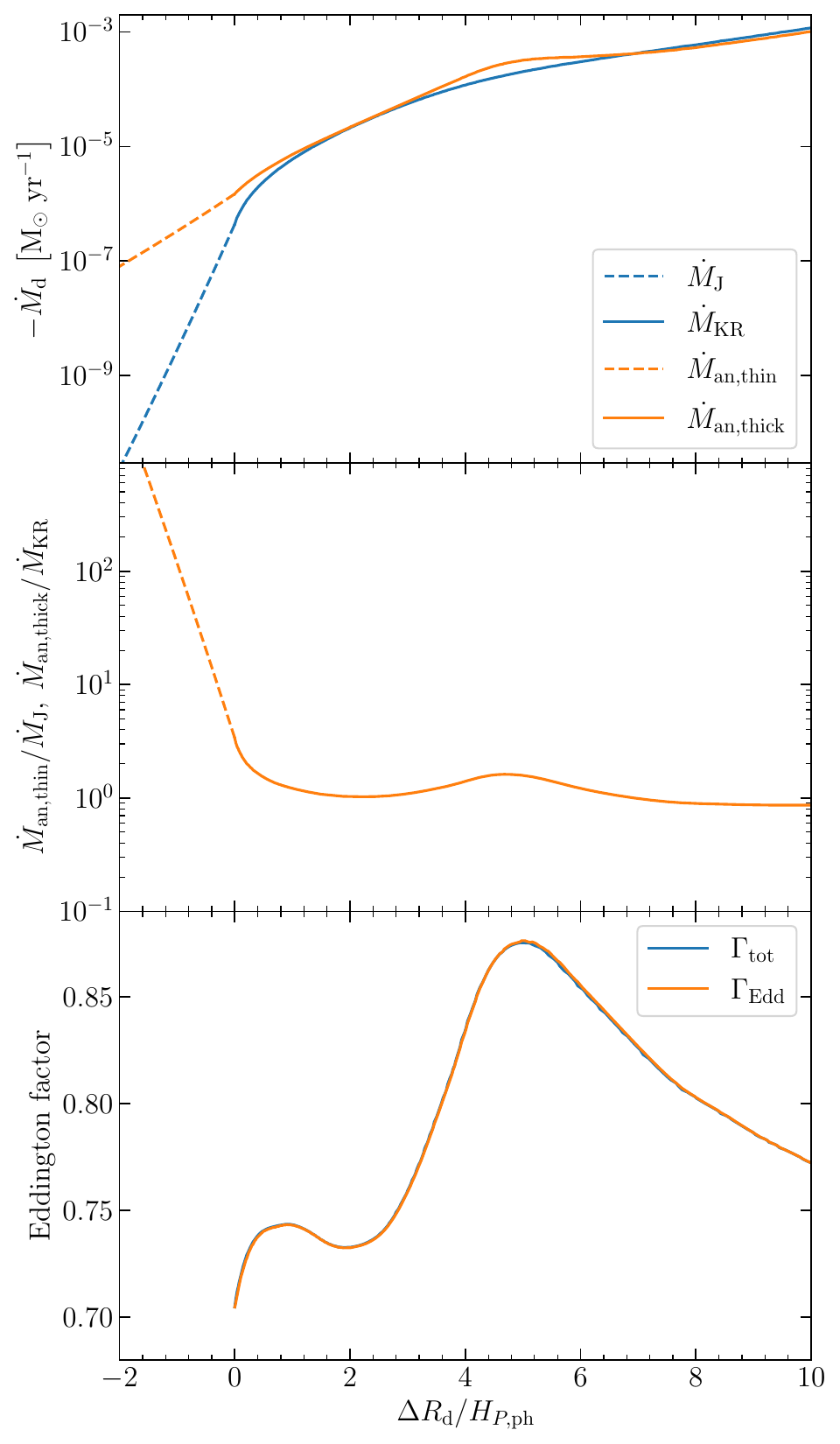}
    \caption{Comparison of our analytical MT prescription $\dot{M}_{\rm an,thin}$ and $\dot{M}_{\rm an,thick}$ with the optically thin prescription introduced by \citet{jackson2017}, $\dot{M}_{\rm J},$ and the optically thick prescription introduced by \citet{kolb1990}, $\dot{M}_{\rm KR}$. We also show the Eddington factors $\Gamma_{\rm tot} = L/L_\text{Edd}$ and $\Gamma_{\rm Edd}$ (equation~\ref{eq:Gamma_Edd}). The case shown is for an isolated originally $40 \rm{M}_\odot$ low-metallicity star in the supergiant stage.}
    \label{fig:Mdot-an-KR-40Msun}
\end{figure}

In Fig.~\ref{fig:Mdot-an-KR-40Msun} we show the comparison of our new analytical estimates $\dot{M}_{\rm an,thin}$ (equation~\ref{eq:dotM_an,thin}) and $\dot{M}_{\rm an,thick}$ (equation~\ref{eq:dotM_an,thick}) with the standard model of optically-thin MT \citep[$\dot{M}_{\rm J}$, equation~\ref{eq:Mdot_J}]{jackson2017} and optically-thick MT \citep[$\dot{M}_{\rm KR}$, equation~\ref{eq:Mdot_KR}]{kolb1990}. Here, we measure the absolute radius excess $\Delta R_\dd$ in the units of the photosphere pressure scale height $H_{P,\rm ph}$ defined as
\begin{equation}
    H_{P,\rm ph} \equiv \frac{P_{\rm ph}}{\rho_{\rm ph}} \frac{R_\dd^2}{G M_\dd},
\end{equation}
where $P_{\rm ph}$ is the photosphere pressure. We see that in the optically thin regime $\dot{M}_{\rm an,thin}$ is greater than $\dot{M}_{\rm J}$ due to the momentum boost from radiation that is included in our expression, and the relative difference is growing exponentially. In contrast, $\dot{M}_{\rm an,thick}$ is similar to $\dot{M}_{\rm KR}$ because the model of \citet{kolb1990} integrates the total pressure (including $P_{\rm rad}$) of the overflowing layers and therefore includes much of the radiation boost. We also confirm that $\dot{M}_{\rm an,thin} \approx \dot{M}_{\rm J}$ and $\dot{M}_{\rm an,thick} \approx \dot{M}_{\rm KR}$ for a gas pressure dominated envelope ($P_{\rm gas} \gg P_{\rm rad}$) on a solar-like donor on the red giant branch, see Appendix~\ref{app:1Msun}.

\subsection{Comparison to previous works}\label{sec:comparison}
There are two main implications for sub-Eddington MT based on our results in this Section. First, for Roche-lobe-underfilling donors (optically thin case), the MT rate increases exponentially with the Eddington factor at the photosphere. This effect is not part of the existing prescriptions, but can be easily included by using equation~(\ref{eq:dotM_an,thin}). Second, for Roche-lobe-overflowing donors, the MT rate is very similar to existing prescriptions because the underlying hydrostatic stellar structure is already taking radiation pressure into account. Subsurface convective regions appear to have only minimal influence on the MT rate. In this regime, we provide a convenient expression (equations~\ref{eq:dotM_an,thick}, \ref{eq:dotM_an,thick,tot}) to evaluate the MT rate without the need to integrate the stellar structure or solve the hydrodynamic equations. The uncertainties arising from using this expression appear similar to the systematic differences between existing MT prescriptions (Fig.~\ref{fig:Mdot-comp}).

We now compare the approach in this paper to our previous work in \citetalias{cehula2023}. For convenience, we summarize the equations from \citetalias{cehula2023} in Appendix~\ref{app:previous_and_other}. The greatest difference between these works comes from the fact that here we assume that all physical quantities are constant along $Q$ and we use a physically motivated expression for $Q$ (equations~\ref{eq:Q}, \ref{eq:dQ_dphi}), where $Q$ is appropriately enlarged by a factor of $1/(1-\Gamma_{\rm Edd})$. In contrast, in \citetalias{cehula2023} we assumed hydrostatic equilibrium in the $yz$-plane together with a polytropic EOS and averaged over the $yz$-plane. As a result, characteristic cross sections corresponding to density and pressure profiles, $Q_\rho$ and $Q_P$, appeared in all equations. Moreover, \citetalias{cehula2023} had an extra term on the right-hand side of the energy equation~(\ref{eq:hydro_gen_en}) that is a consequence of the average of the Roche potential $\phi_{\rm R}$ in the $yz$ plane. The effects of this term can be understood by realizing that if the gas is farther from the $x$ axis, its potential is greater. Consequently, if the cross section shrinks, the potential energy becomes lower, and this energy difference is used to heat the gas. Thus, our original \citetalias{cehula2023} model is not strictly adiabatic but is adiabatic in the sense that there is no energy exchange between the gas and radiation. 

In the present paper, we define two limiting cases: the small-$v$ case that is inspired by the hydrostatic case (Section~\ref{sec:small_v}) and the adiabatic case with no energy exchange between gas and radiation and no extra term on the right-hand side of the energy equation (Section~\ref{sec:adiabatic}). The adiabatic case behaves similarly to the original ``adiabatic'' \citetalias{cehula2023} model with the difference that the adiabatic case gives a steeper temperature gradient resulting in a lower temperature at L1 and thus lower critical velocity $v_{\rm crit}$ and a lower value of the MT rate. Also, if $x_0$ approaches $x_1$,  the two cases obtain different $Q$ and $v_\text{crit}$ and consequently also different MT rates. Specifically, the adiabatic case obtains $Q \propto P_{\rm gas}/\rho/(1-\Gamma_{\rm Edd}) \propto P/\rho$ and $v_{\rm crit}^2 = P_{\rm gas}/\rho$, while our original model had $Q \propto c_{T, \rm gas}^2 \propto P_{\rm gas}/\rho$ and $v_{\rm crit}^2 \sim P/\rho$. Hence, the original model obtains a lower value of the MT rate even in this limiting case as $-\dot{M}_{\rm d} \propto v_{\rm crit} Q$. The small-$v$ case and the adiabatic case obtain the same MT rate in the limit of $x_0 \to x_1$ because they predict the same $Q$ and $v_\text{crit}$. This behavior is illustrated in Figs.~\ref{fig:prof-comp-m7} and \ref{fig:Mdot-R0-comp-m7}.

\section{Super-Eddington mass transfer}\label{sec:superedd}

We now analyze situations where the star exceeds the Eddington limit interior of the L1 point. We could still solve our equations with $x_0$ in a sub-Eddington region and define $x_1$ as the point where $\Gamma_\text{Edd} = 1$. However, super-Eddington flux can often be carried out by convection without any outflows. Outflows may only develop when convective efficiency diminishes and $v_\text{conv}$ approaches the sound speed. In principle, our equations could be extended to include convective energy transport using, for example, the formalism of \citet{kuhfuss1986}, where the outflow should naturally develop once convection is unable to carry the flux. However, studies of single nonrotating super-Eddington stars have shown that multidimensional effects fundamentally alter the structure of the relevant regions \citep[e.g.,][]{shaviv01a,owocki2004,jiang2015,jiang2018}. This suggests that further refinements of our 1D model may be of limited value. Instead, we revisit some of the arguments constraining super-Eddington outflows in spherical symmetry and find that proximity to the L1 point may relax some of them. While our results are promising with respect to the potential for a super-Eddington MT boost, determining whether such outflows actually occur will ultimately require multi-dimensional time-dependent simulations.

\subsection{Analytic theory}\label{sec:superedd_analytic}
We need to consider three issues determining the viability of a super-Eddington boost to MT: \emph{(i)} efficiency of convection for transferring the energy flux, \emph{(ii)} properties of the super-Eddington MT boost such as the MT rate, and \emph{(iii)} whether the flow satisfies the limits of photon tiring. The discussion of the first issue does not depend on the geometry and we can simply assume that convection becomes unable to carry the super-Eddington flux once the required velocities of convective eddies $v_\text{conv}$ exceed the local sound speed, or alternatively when $L_\text{d} \gtrsim 4\pi r^2 \rho c_{T,\text{gas}}^3$.

For the discussion of the two remaining issues, we first outline the theory of super-Eddington outflows from spherically-symmetric stars presented by \citet{owocki2004} and later revisited by \citet{quataert16} and \citet{owocki2017}, and then modify the results to the context of the L1 point. For $\Gamma_\text{tot} = L/L_\text{Edd} > 1$ and inefficient convection, the entire luminosity can be carried by an outflow with velocity $v_\text{out} \gtrsim c_{T,\text{gas}}$ satisfying
\begin{equation}
    L_\text{d} \sim 4 \pi R_\text{d}^2 \rho v_\text{out}^3,
    \label{eq:L_vout}
\end{equation}
where we assume that the launching region with density $\rho$ is located close to the surface $R_\text{d}$. This implies a spherically symmetric mass-loss rate of
\begin{equation}
    \dot{M}_{\rm out} = 4 \pi R_{\rm d}^2 \rho v_{\rm out} \sim \frac{L_{\rm d}}{v_{\rm out}^2}.
\end{equation}
The outflow velocity $\vout$ depends on $\Gamma_\text{tot}$, but part of the luminosity can still be transported by radiation, making the estimate of $\vout$ in equation~(\ref{eq:L_vout}) effectively an upper limit.

The super-Eddington outflow, however, should not exceed the photon-tiring limit
\begin{equation}
    \dot{M}_{\rm tir} \equiv \alpha_{\rm tir} \frac{L_{\rm d}}{{G M_{\rm d}}/{R_{\rm d}}} = \alpha_{\rm tir} \frac{2 L_{\rm d}}{v_{\rm esc}^2},
\end{equation}
where $\alpha_\text{tir} = 1 - 1/\Gamma_\text{tot}$ is the efficiency parameter of photon tiring \citep{owocki2017}. If $\dot{M}_\text{out} \gtrsim \dot{M}_\text{tir}$, the outflow should just stall. The condition for an outflow to form is thus
\begin{equation}
    \beta_{\rm SE} \equiv \frac{1}{2 \alpha_{\rm tir}} \frac{v_{\rm esc}^2}{v_{\rm out}^2} \lesssim 1.
\end{equation}
Since $v_\text{esc} \gg c_{T,\text{gas}} \sim v_\text{out}$ and $\alpha_{\rm tir} < 1$, this condition is not satisfied and the spherically-symmetric outflow should always stall. 

\begin{figure*}
    \centering
    \includegraphics[width=0.99\textwidth]{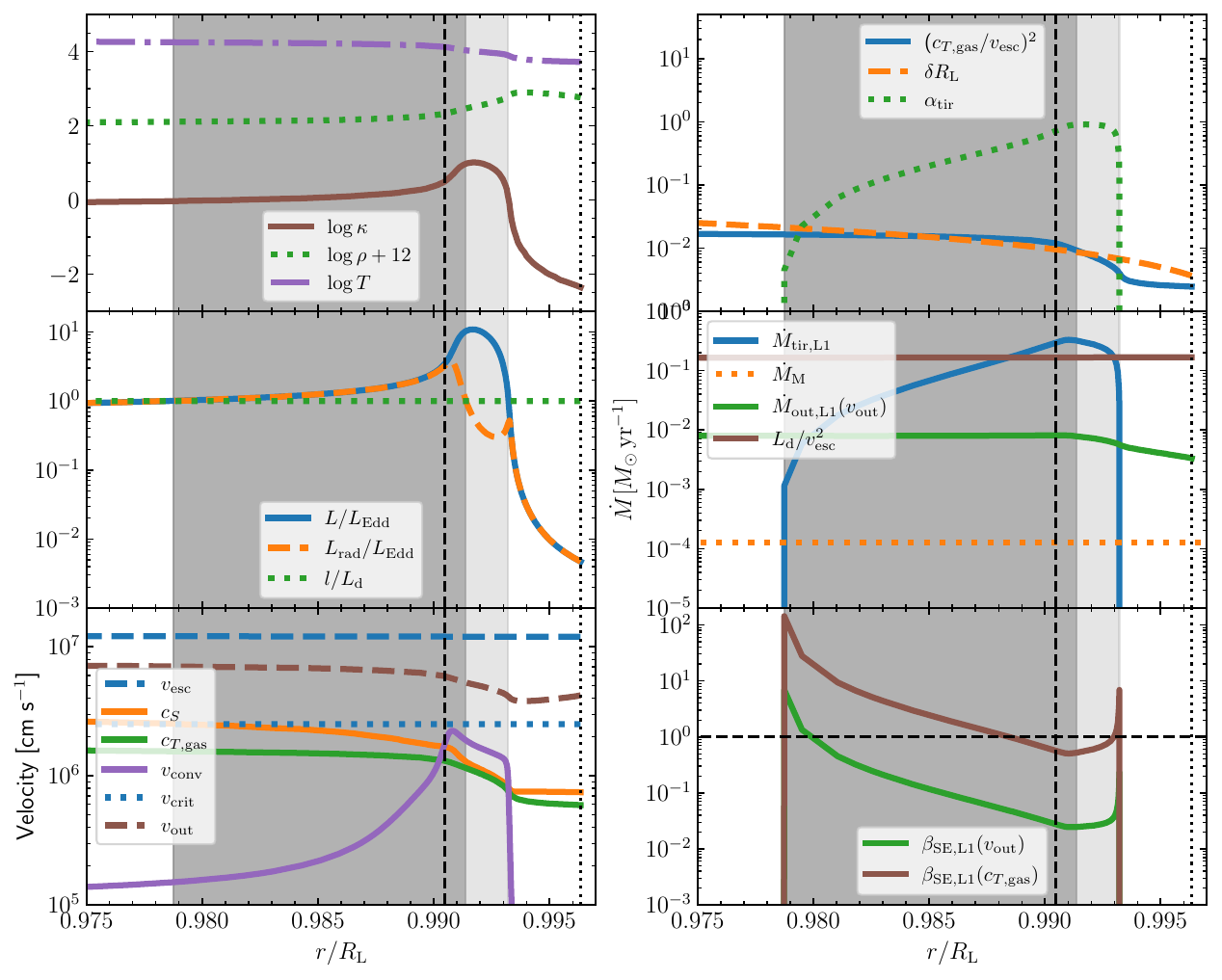}
    \caption{Radial profiles of key quantities for an initially $30\,\rm{M}_\odot$ low-metallicity star during the super-Eddington MT boost at $t=6.972$\,Myr, when the donor has $M_\text{d} = 28.57\,\msun$, $T_\text{eff} = 5255$\,K, $L_\text{d} = 3.96 \times 10^5\,\rm{L}_\odot$, and $R_\text{d} = 759\,\rm{R}_\odot$. Upper left panel shows profiles of density, temperature, and opacity. Middle left panel shows Eddington ratios of total ($L$) and radiative ($L_\text{rad}$) luminosities and relative luminosity profile $l/L_\text{d}$. Bottom left panel shows values of escape velocity $\vesc$, sound speeds $c_S$ and $\ctgas$, convective velocity reported by the MLT modules in MESA $v_{\rm conv}$, outflow velocity $v_\text{out}$ (equation~\ref{eq:L_vout}), and the critical velocity $v_\text{crit}$ of \citet{quataert16}. Upper right panel displays the estimate for L1 nozzle area $\ctgas^2/\vesc^2$, relative distance to the L1 point $\delta R_\text{L}$, and efficiency parameter of photon tiring $\alpha_\text{tir}$. Middle right panel shows estimates of L1 MT rates based on super-Eddington boosted MT $\dot{M}_\text{out,L1}$ (equation~\ref{eq:mdot_outL1}), a more pessimistic estimate with $v_\text{out}$ replaced with $\ctgas$, $L_{\rm d}/ v_{\rm esc}^2$, and the prediction of \citet{marchant2021} $\dot{M}_\text{M}$. In addition, we show also the photon tiring limit $\dot{M}_\text{tir,L1}$ (equation~\ref{eq:mdot_tirL1}). Bottom right panel shows the condition for super-Eddington boost $\beta_\text{SE,L1}$ evaluated using $v_\text{out}$ or $\ctgas$ (equation~\ref{eq:beta_seL1}). In all panels, light and dark shading marks super-Eddington regions based on $L$ and $L_\text{rad}$, vertical dashed line shows the interior most point where $v_\text{conv} > c_S$, and the vertical dotted line indicates the position of the photosphere. \label{fig:superedd}}
\end{figure*}

Models based on the porosity length formalism suggest that the maximum achievable mass loss rate is proportional to the gravitational scale height \citep[e.g.][]{owocki2004}, and thus inversely proportional to effective gravity $g_{\rm eff} \propto \nabla \phi_{\rm R}$ in a Roche model, where $\nabla \phi_{\rm R} \to 0$ close to L1. Consequently, we assume that if an outflow occurs from a star nearly filling its Roche lobe and possessing a super-Eddington, convectively inefficient subsurface zone, it will primarily flow through the nozzle surrounding L1 at a rate $\mdotoutLone$. The detailed steps needed for the modifications from spherical symmetry to the vicinity of L1 are described in Appendix~\ref{app:dotM_pt,L1}. In summary, we take into account the limited extent of the nozzle by multiplying $\mdot$ by $Q/(4\pi R_\text{d}^2) \sim c_{T,\text{gas}}^2/v_\text{esc}^2$, and differences in potential proportional to $\vesc^2$ are multiplied by $\deltaRL$. These modifications give
\begin{align}
        \mdotoutLone &\equiv \frac{Q}{4 \pi R_{\rm d}^2} \dot{M}_{\rm out} &&= \frac{c_{T, \rm gas}^2}{v_{\rm esc}^2} \frac{L_{\rm d}}{v_{\rm out}^2},\label{eq:mdot_outL1}\\
        \dot{M}_{\rm tir, L1} &\equiv \alpha_{\rm tir} \frac{L_{\rm N}}{\bar{\phi}(R_{\rm L}) - \bar{\phi} (R_{\rm SE})} &&= \alpha_{\rm tir} \frac{c_{T, \rm gas}^2}{v_{\rm esc}^2 \delta R_{\rm L}} \frac{2 L_{\rm d}}{v_{\rm esc}^2},\label{eq:mdot_tirL1}
\end{align}
where $R_{\rm SE}$ is a radius inside the donor where $\Gamma_{\rm Edd} > 1$ and the condition for the outflow not to stall reads
\begin{equation}
    \beta_{\rm SE, L1} = \frac{1}{2 \alpha_{\rm tir}} \frac{\vesc^2}{\vout^2} \deltaRL = \beta_{\rm SE} \deltaRL \lesssim 1.\label{eq:beta_seL1}
\end{equation}
Since $\delta R_\text{L}$ can be made arbitrarily small by positioning the super-Eddington layer very close to the Roche lobe, this modified condition can be satisfied and the super-Eddington outflow will not be limited by photon tiring. In such a case, a super-Eddington outflow with velocity $v_{\rm out}$ and MT rate $- \dot{M}_{\rm d} \sim \dot{M}_{\rm out, L1}$ forms through the L1 point. Let us note here that the outflow rate $\dot{M}_{\rm out, L1}$ does not depend on the distance from the L1 point, but only on the properties of the donor. The condition (\ref{eq:beta_seL1}) is satisfied close to the L1 point because the photon-tiring limit $\mdotoutLone$ increases as the super-Eddington layer approaches the L1 point.

\citet{quataert16} presented a different formulation of the photon tiring limit. They argued that equation~(\ref{eq:L_vout}) is valid when the critical velocity of the outflow, $\vcrit \sim (GL_\text{d})^{1/5}$, exceeds the local escape velocity modified by a factor taking into account the structure of the surrounding envelope. In the vicinity of the L1 point, their condition reads $\vcrit \gtrsim \vesc \deltaRL^{1/2}$. We show this condition in the figures that illustrate the super-Eddington MT boost, but we do not find significant differences in the outcomes when compared to equation~(\ref{eq:beta_seL1}).

\subsection{Application to a realistic star}\label{sec:superedd_app}

The higher the photon-tiring limit $\dot{M}_{\rm tir,L1}$ (equation \ref{eq:mdot_tirL1}), the easier it is to form an outflow through the L1 point. Therefore, we want the Eddington factor $\Gamma_{\rm Edd}$ to be as high as possible in the super-Eddington layer. The maximum Eddington factor inside stars was investigated by \citet{sanyal2015}, who analyzed massive stars at LMC metallicity. They found that the local Eddington factor can exceed seven due to the hydrogen opacity bump near the photosphere in stars with effective temperatures around 5500 K (see their fig. 3). Fortunately, a similar situation arises in the $30\,\mathrm{M}_\odot$ low-metallicity donor studied by \citet{marchant2021}.

Thus, to illustrate the super-Eddington MT boost we use the same $30\,\rm{M}_\odot$ low-metallicity donor as in Sec.~\ref{sec:30Msun} but this time in a binary with an initial $4.5\,\rm{M}_\odot$ black hole and initial orbital period of $1000$ days. As before, we recalculated the evolution using the full MT prescription $\dot{M}_{\rm{M}}$ and the same MESA version as in \citet{marchant2021}. Our first goal is to investigate how to evaluate the conditions for a super-Eddington MT boost on realistic stellar profiles. In Fig.~\ref{fig:superedd}, we show radial profiles of a number of relevant quantities near the outer edge of the stellar envelope at a moment, when the donor still underfills its Roche lobe. Donor radius is $R_{\rm d} = 0.9963 R_{\rm L}$ (vertical dotted line in Fig.~\ref{fig:superedd}), which means that the relative radius excess is $\delta R_{\rm d} = -0.00365$. We see that the super-Eddington layer reaches $\Gamma_\text{tot} \gtrsim 10$, occurs in the hydrogen recombination zone, and is accompanied by a density inversion (upper left panel). The rapid decline of $\Gamma_{\rm tot}$ near the photosphere is caused by the sharp drop in opacity (upper left panel), while the luminosity $L$ remains constant throughout the donor’s envelope (middle left panel). We find that simple estimates of the convective energy flux differ sufficiently from the results of mixing length theory (MLT) of convection implemented in MESA and are not useful. Instead, we diagnose the inefficiency of convection by comparing the $v_\text{conv}$ reported by MLT with the local sound speed (lower left panel), finding that convection is inefficient in a region $0.990 \lesssim r/R_\text{L} \lesssim 0.993$. This result is insensitive to the use of $c_S$ or $c_{T, \text{gas}}$ for comparison purposes. The lower left panel of Fig.~\ref{fig:superedd} also shows $v_\text{crit}$. Near the L1 point, we find that the relevant quantities for the \citet{quataert16} version of the photon tiring are $v_\text{crit} \approx 25\,$km\,s$^{-1}$, $v_\text{esc} \approx 120\,$km\,s$^{-1}$, and $\delta R_\text{L}^{1/2} \approx 0.1$ implying that the requirement is satisfied and equation~(\ref{eq:L_vout}) can be used to find $v_\text{out}$. Our results indicate that $v_\text{out}$ exceeds the local sound speed by a factor of 3--4, but remains smaller than $v_\text{esc}$. The middle right panel of Fig.~\ref{fig:superedd} displays $\dot{M}$ near L1 corresponding to the MT flow and the photon tiring limit. The shape of $\dot{M}_\text{tir}$ is mostly determined by $\alpha_\text{tir}$ (upper right panel). We find that $\dot{M}_\text{out,L1} < \dot{M}_\text{tir,L1}$ and correspondingly $\beta_\text{SE,L1} < 1$ in a significant range of $r$ (bottom right panel). Since there is some ambiguity about the correct value of $v_\text{out}$, we also show an estimate of $\dot{M} = (c_{T,\text{gas}}/v_\text{esc})^2 L_{\rm d} / c_{T, \rm gas}^2 = L_{\rm d}/\vesc^2$ and modified $\beta_\text{SE,L1}(\ctgas)$, where $\vout$ is replaced with $\ctgas$ in equation~(\ref{eq:beta_seL1}). Even in this more pessimistic case, the implied MT rate does not exceed the photon tiring limit. While in this case the outflow condition is satisfied later in the evolution (only about 700 years before Roche-lobe overflow), the implied MT rates are higher leading to a higher total mass lost. 

\begin{figure}
    \centering
    \includegraphics[width=0.99\columnwidth]{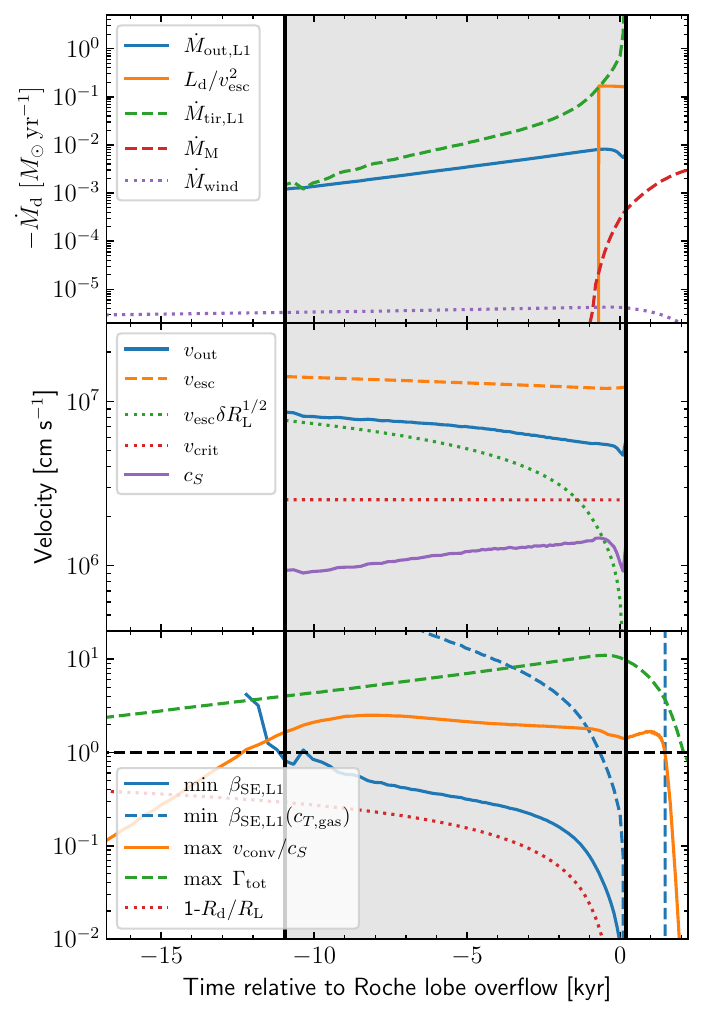}
    \caption{Time evolution of donor properties illustrating super-Eddington boosted MT on an initially $30\,\msun$ star in a binary with a $4.5\,\msun$ black hole. Top panel shows estimates of L1 MT rates $\dot{M}_\text{out,L1}$ (equation~\ref{eq:mdot_outL1}), $\dot{M}_\text{M}$ \citep{marchant2021}, and a more pessimistic estimate for super-Eddington MT $\mdot = L_{\rm d}/\vesc^2$, as well as photon tiring limit $\dot{M}_\text{tir,L1}$ and wind mass loss from the donor $\dot{M}_\text{wind}$. Middle panel shows the evolution of relevant velocities $v_\text{out}$, $\vesc$, $v_\text{crit}$, and $c_S$. Bottom panel shows time evolution of the minimum of $\beta_\text{SE,L1}$, maximum of $v_\text{conv}/c_S$, and maximum of $\Gamma_\text{tot} = L/L_{\rm Edd}$ evaluated over the subsurface region that could drive super-Eddington MT, and the relative Roche-lobe-filling factor of the star $1-R_\text{d}/R_\text{L}$.\label{fig:timeev}}
\end{figure}

To summarize Fig.~\ref{fig:superedd}, we find that at this point in evolution the donor is underfilling its Roche lobe, but the super-Eddington flux in the subsurface layer probably leads to MT through L1 with a rate of around $10^{-2}\,\msunyr$, which is significantly lower than the photon tiring limit near L1 but also significantly higher than the MT rate of about $10^{-4}\,\msunyr$ calculated using the traditional approach \citep{marchant2021}. 

We now proceed to evaluate the time dependence of the super-Eddington MT boost and its importance for binary evolution by post-processing the evolution calculated using the traditional MT model without the super-Eddington boost. This serves to illustrate the possible outcomes rather than provide quantitative predictions, which would require a self-consistent evolutionary calculation. In Fig.~\ref{fig:timeev}, we show the time evolution of relevant quantities a few thousand years before and after Roche-lobe overflow, which occurs at about $6.973$\,Myr and which we set as $t=0$. We find that the donor with the super-Eddington subsurface layer expands and the convection efficiency decreases until about $t=-12$\,kyr when $v_\text{conv} > c_S$. Shortly after, at $t=-11\,$kyr, $\beta_\text{SE,L1}$ decreases below unity, and the donor should begin experiencing super-Eddington boosted MT. The implied MT rates of $10^{-3}$ to $10^{-2}$\,\msunyr\ dramatically exceed the traditional optically thin MT rates or the donor's mass loss due to stellar winds \citep{marchant2021}. Since on surface $\Gamma_\text{tot} \lesssim 0.1$, the Eddington boost to the optically thin MT rate discussed in Section~\ref{sec:iso} is not strong. For the final $1$\,kyr before Roche-lobe overflow, the traditionally calculated MT rate quickly increases and approaches $10^{-3}$\,\msunyr. At $t=0$, the super-Eddington MT boost formally ends because the driving region moves outside the L1 point.

By integrating $\dot{M}_\text{d}$, we find that the total mass lost from the donor due to the super-Eddington MT exceeds $10\,\msun$. We obtain even higher numbers for the more pessimistic case when $v_\text{out} = \ctgas$ or similar numbers even when we restrict the super-Eddington MT to the final $1$\,kyr before Roche-lobe overflow when $v_\text{crit} > \vesc \delta R_\text{L}^{1/2}$. These numbers are clearly an artifact of not evolving the binary self-consistently. We performed preliminary MESA simulations with the super-Eddington boost included (Hada\v{c}, Cehula, Pejcha, in preparation) and found that the star responds to pre-Roche overflow mass loss similarly to standard mass transfer, leading to overall similar evolution, however, the timing and observable signatures of the mass loss are different.

\section{Summary and discussion}\label{sec:discussion}

In this paper, we investigated the effect of radiation pressure on MT in binary stars. Starting from 3D  time-steady radiation hydrodynamic equations in the flux-limited diffusion approximation (equations~\ref{eq:3D}), we constructed 1D equations describing the flow near the L1 nozzle (equations~\ref{eq:1D}). Using the von Zeipel theorem (Section~\ref{sec:von_zeipel}), we defined several simplified cases in the limit of small velocities and optically thick regions (Section~\ref{sec:limiting}), where the flow is parameterized by the Eddington parameter $\Gamma_\text{Edd}$ and the critical point occurs either at the L1 point or at $\Gamma_\text{Edd} = 1$. For sub-Eddington donors, $\Gamma_\text{Edd} < 1$, we were able to find several analytic estimates and algebraic solutions (Section~\ref{sec:analytic} and \ref{sec:algebraic}) as well as numerically solve the equations in the context of massive low-metallicity donor (Section~\ref{sec:real_EOS}). For donors with a super-Eddington layer below the surface, we reformulated the limitations for outflows posed by convective efficiency and photon tiring and applied the theory to a binary evolution trajectory (Section~\ref{sec:superedd}). Our results can be summarized as follows.
\begin{itemize}
\item If the donor envelope is dominated by gas pressure, $P_{\rm gas} \gg P_{\rm rad}$ or equivalently $\Gamma_{\rm Edd} \ll 1$, our model predicts a MT rate comparable to standard models by \citet{jackson2017} and \citet{kolb1990} in the optically thin and thick regimes, respectively. This is illustrated for a solar-like donor on the red giant branch in Fig.~\ref{fig:Mdot-an-KR-1Msun}.  
\item If the donor envelope is dominated by radiation pressure, $P_{\rm rad} \gg P_{\rm gas}$ and $\Gamma_{\rm Edd} \lesssim 1$, our optically thick rates are comparable to predictions of \citet{kolb1990} and later works based on similar assumptions (Fig.~\ref{fig:Mdot-an-KR-40Msun}). The reason is that radiation pressure is already accounted for in the hydrostatic stellar structure used as input in the existing models. However, for Roche-lobe-underfilling donors, our optically thin MT rates are significantly higher than the predictions of previous models of \citet{ritter1988} or \citet{jackson2017} (Fig.~\ref{fig:Mdot-an-KR-40Msun}). The increase in the MT rate is exponential in $1-\Gamma_\text{Edd}$ and the effect can be included by a simple modification of existing prescriptions (equation~\ref{eq:dotM_an,thin}). 
\item We derived an analytic approximation for the optically thick MT rate, which is based only on the quantities at the L1 point (equation~\ref{eq:dotM_an,thick}). We verified that this approximation differs from numerical solutions by factors similar to the spread among existing MT models (Figs.~\ref{fig:Mdot-R0-comp-m7} and \ref{fig:Mdot-comp}).
\item Super-Eddington convectively inefficient subsurface layers in single non-rotating donors lead to envelope inflation and density inversions, but should not develop super-Eddington outflows, because the implied mass-loss rates exceed the photon tiring limit \citep{owocki2004}. In a binary system,  the L1 point effectively brings the gravitational potential infinity much closer to the super-Eddington layer and effectively increases the photon tiring limit (equation~\ref{eq:beta_seL1}). As a result, a super-Eddington-boosted MT could occur (Fig.~\ref{fig:superedd}). By postprocessing a binary evolutionary trajectory, we proposed that super-Eddington-boosted MT can occur before the donor overfills its Roche lobe and that MT rates reach $\gtrsim 10^{-2}\,\msunyr$ (Fig.~\ref{fig:timeev}). Consequently, this new mode of MT can fundamentally alter the evolutionary trajectories of massive binaries.
\end{itemize}

\subsection{Implications}\label{sec:implications}
We have identified two new effects of radiation pressure on MT: the exponential increase in the optically thin MT rate for donors with a high but sub-Eddington surface $\Gamma_\text{Edd}$ and the circumvention of the photon tiring limit in donors possessing super-Eddington convectively inefficient subsurface layers. The importance of the increased optically thin MT rate for binary evolution is likely limited, because although the rates are much higher than traditionally assumed, they are still significantly lower than MT rates for Roche-lobe-overflowing donors or the mass lost from a massive donor by stellar winds. Moreover, the details of MT via the L1 point or the precise value of the Roche-lobe radius are not crucial for binary evolution, provided that the MT is stable and the long-term evolution is driven by other factors such as the nuclear or thermal timescale of the donor, as already pointed out by \citet{paczynski1971}. In such cases, it is more important to consider mass and angular momentum loss from the binary.

Indeed, the enhanced optically thin rate could be important for donors undergoing wind Roche-lobe overflow and for the stability of massive contact binaries, which are already close to overflowing their outer Lagrangian points (L2 and L3). If substantial mass is lost through L2/L3 via an optically thin outflow analogous to a similar process near L1, it would carry away angular momentum, causing the binary to shrink. If this effect is strong enough, it could destabilize the contact binary, potentially leading to a runaway merger with consequences similar to what was observed in a low-mass merger V1309~Sco \citep[e.g.,][]{tylenda2011,pejcha2014,pejcha2017}. To assess this impact, future simulations should incorporate our MT model at the L2/L3 points.

The consequences of the potential super-Eddington MT boost in donors with a convectively inefficient subsurface layer are far-reaching. In Fig.~\ref{fig:timeev}, we only illustrate the possible effects by post-processing a binary evolution trajectory. Although the quantitative details are unreliable, we can still deduce general features that are likely valid. First, the super-Eddington MT boost can occur tens of kyr before the donor formally overflows its Roche lobe. The initial growth timescale of $\dot{M}_\text{out,L1}$ is difficult to assess from the post-processed run, but the behavior of $\beta_\text{SE,L1}$ in Fig.~\ref{fig:timeev} suggests that it is probably around a thousand years and possibly even less. The super-Eddington boost in Fig.~\ref{fig:timeev} formally ends when the driving region moves beyond the L1 point, which becomes the innermost critical point of the outflow and the physical conditions at this point will determine $\dot{M}_\text{d}$ (equation~\ref{eq:3D_vel_id}). It is reassuring that at $t=0$, $\dot{M}_\text{M}$ is not too different from $\dot{M}_\text{out,L1}$. At $t>0$, the effects of radiation pressure will likely affect the outflow velocity, which could become noticeably higher than the sound speed at L1. 

The total amount of mass transferred during the super-Eddington boost depends on how the donor structure responds to this type of mass loss. We can imagine two scenarios. First, the super-Eddington MT simply adds to the traditional MT rate. The effective outcome of this could be simply that the Roche-lobe overflow occurs earlier than traditionally predicted, but the ultimate fate of the binary remains unaffected. Alternatively, if the mass transfer is already unstable and a hydrostatic structure can no longer be maintained in the outer layers, this additional boost could further accelerate the instability. 

Second, the high $\dot{M}_\text{out,L1}$ induces a strong response in the donor and the super-Eddington convectively inefficient layer disappears and the donor shrinks. The super-Eddington layer could reconstitute again, leading to expansion and repeating of the super-Eddington-boosted MT. In such a case, the donor might completely avoid overfilling its Roche lobe, and the binary evolution would be significantly altered. However, it is not clear how much different the ultimate fate of the binary would be.

This last scenario motivates further speculation on the cause of LBV outbursts. LBVs are hot, highly luminous massive stars that undergo large, quasi-periodic variations in brightness, radius, and photospheric temperature on timescales of years to decades, often accompanied by episodes of enhanced mass loss \citep{smith2014}. The connection between these eruptions and super-Eddington outflows has a long history \citep[e.g.,][]{smith_owocki06}, but it has not been clear what triggers the enhanced super-Eddington state. 

A single-star scenario was recently proposed by \citet{grassitelli2021}. In addition to an inflated envelope close to the Eddington limit, they required two additional conditions to trigger the bursting activity: first, a temperature range in which decreasing opacities do not lead to an accelerating outflow; and second, a mass-loss rate that increases with decreasing temperature.

Our results in Fig.~\ref{fig:timeev} might lead to a binary scenario for LBV outbursts. In addition to a super-Eddington, convectively inefficient subsurface layer, we also require this layer to be sufficiently close to the L1 point. When it is far from L1 and unable to drive an outflow due to photon tiring, it becomes subject to instabilities leading to up and down flows as well as small eruptions \citep[e.g.,][]{owocki2004,jiang2015,jiang2018}. Once the donor slowly expands sufficiently close to the L1 point, the photon tiring limit is relaxed and the donor is able to drive an outflow through the L1 nozzle. Envelope instabilities are likely to make this process manifest as a series of distinct irregular bursts rather than a smooth outflow. The expected mass-transfer/outflow rates are consistent with what was inferred for LBV outbursts \citep{smith2014}. Depending on how the donor responds to this type of mass loss, there might be multiple phases of activity separated by inactive periods during which the driving region in the donor reconstitutes. In general, the donor might lose up to several $\msun$ of material before the conditions for the outflow are no longer satisfied.

At first sight, such a scenario might seem extremely unlikely to occur. In this paragraph, we argue that this is not necessarily the case. If we evolve the $30\,\mathrm{M}_\odot$ low-metallicity donor from Sec.~\ref{sec:superedd_app} as a single star, the super-Eddington and convectively inefficient subsurface layer is present for $R_{\rm d}/R_\odot$ between roughly 540 and 910. Assuming that the donor is in a binary, Keplerian orbit, and adopting the simple approximation for the Roche-lobe radius from \citet{paczynski1971}, $R_{\rm L}/a = 0.46 [q/(1+q)]^{1/3}$, we obtain the following expression for the binary orbital period: $P_{\rm orb} = 20\sqrt{R_{\rm L}^3/G M_{\rm d}} = 2.3~\mathrm{yr}~R_{\rm L,540}^{3/2} M_{\rm d,30}^{-1/2}$, where $R_{\rm L,540} = R_{\rm L}/540\,\rm{R}_\odot$ and $M_{\rm d,30} = M_{\rm d}/30\,\rm{M}_\odot$. This expression does not depend on the binary mass ratio. Thus, if the $30\,\mathrm{M}_\odot$ low-metallicity donor resides in a binary with an orbital period between roughly 2.3 and 5.1~years, it should overflow its Roche lobe while possessing a super-Eddington, convectively inefficient subsurface layer, and a super-Eddington MT boost could occur. Given that most massive stars are in binaries \citep[e.g.][]{sana2012}, these numbers already suggest that such a scenario might not be extremely rare.

The diversity of outcomes of this binary scenario for LBV outbursts can be further increased by considering super-Eddington layers due to hydrogen, helium, or iron ionization, accretion onto the companion, which might be a compact object able to accelerate part of the material to high velocities, or eccentric passages of the companion, where the effective L1 point passes close to the surface only temporarily. All of these issues are nontrivially mapped on the properties of the initial conditions of binary components and their orbit. However, this scenario cannot possibly explain all LBV outbursts. For example, since the driving mechanism ultimate relies on stellar photons, the kinetic energy of the ejected material is limited by the energy radiated by the star, which is violated in the brightest LBV eruptions such as $\eta$~Car \citep{smith2011}.

\subsection{Limitations and future work}\label{sec:limit_and_future}
We now discuss several limitations of our work and how we can improve our results in the future. The key assumption in our model is the von Zeipel theorem. Although this theorem is valid in stellar interiors and for slowly rotating stars, it becomes less reliable near the stellar surface and Lagrangian points. Moreover, it is inherently inconsistent. In a stationary system where the radiative flux governs energy transport, energy conservation dictates a divergence-free flux (equations~\ref{eq:3D_4}, \ref{eq:3D_5}). However, the assumption of proportionality between radiative flux and effective gravity (equation~\ref{eq:von_Zeipel}) implies the existence of the Poisson equation in the donor envelope, $\Delta \phi_{\rm R} = 0$, in regions where $L(\phi_{\rm R}) \approx \rm{const.}$ and $M(\phi_{\rm R}) \approx \rm{const}$. However, in reality, $\Delta \phi_{\rm R} = -2 \omega^2$, where $\omega$ is the binary orbital frequency. This inconsistency was resolved in 2D by \citet{espinosa-lara2011}, which suggests a promising avenue for adopting their approach along the binary axis and for further generalization to 3D.

Another simplification of our 1D approach is the neglect of circulation (due to the one-dimensionality of the method) and, more importantly, the neglect of the Coriolis force. Recently, \citet{ryu2025} performed 3D hydrodynamical simulations of MT through the L1 point, including the Coriolis force and assuming an ideal-gas EOS. They found that the Coriolis force affects the stream morphology by reducing the stream’s cross-section and shifting its origin to the donor’s trailing side. However, the resulting MT rates were only mildly suppressed relative to analytic predictions, within a factor of 2 at the L1 point. Without the Coriolis force, their results resemble our nozzle picture and agree with analytic prescriptions to within 5 percent. This suggests that including the Coriolis force in our model could mildly reduce the MT rate (by up to a factor of a few), but this effect should remain subdominant compared to, for example, the super-Eddington MT boost discussed in this work.

Our model for the super-Eddington MT boost in Section~\ref{sec:superedd_analytic} assumes that the material will flow around L1 once the photon tiring limit becomes relaxed there. However, this is only a necessary but not sufficient condition. Instead, it is possible that lateral structuring or ``porosity'' will reduce the effective coupling between radiation and matter similar to what was previously argued for spherical stars \citep[e.g.,][]{shaviv01b,owocki2004}, leading to a time-dependent but still nearly hydrostatic structure \citep{jiang2015,jiang2018}.

There are several ways to further improve our model and understanding of MT in binaries in general. As a first step, 1D binary evolution models that include modifications to the optically thin MT and some prescription for the possible super-Eddington boost would be useful to investigate self-consistent response of the binary to these new modes of MT and to assess the effect on the outcomes of binary evolution and to connect to phenomena like LBV outbursts. However, the model of super-Eddington boost presented in Section~\ref{sec:superedd} depends on several multidimensional radiation-hydrodynamic phenomena: von Zeipel theorem in Roche geometry, photon porosity of radiation pressure-dominated layers, inefficient convection, and the Coriolis force. Although effective 1D theories exist for each of these phenomena separately, capturing the mutual non-linear interactions in the context of some new 1D theory of super-Eddington MT boost is unlikely to be successful. Instead, future multidimensional, time-dependent radiation hydrodynamic simulations of MT through L1 are essential to test the validity of the simple model presented here and to further clarify the flow geometry, MT rate, and response to the super-Eddington condition \citep[e.g.,][]{jiang2015,jiang2018,dickson2024, ryu2025}. Given the difficulty of capturing 3D phenomena with 1D models, the relatively small but noticeable differences in the existing 1D MT prescriptions (Fig.~\ref{fig:Mdot-comp}) suggest that using a simple analytic prescription such as the one presented here in Section~\ref{sec:analytic} might do as good a job as some more complicated 1D models.

\section*{Acknowledgements}

JC thanks Damien Gagnier for useful comments. This research was supported by Czech Science Foundation Grant No. 24-11023S. JC also acknowledges support by the ETAg grant PRG2159 during manuscript revision.

%%%%%%%%%%%%%%%%%%%%%%%%%%%%%%%%%%%%%%%%%%%%%%%%%%
\section*{Data Availability}

The data underlying this article will be shared on reasonable request to the corresponding author.



\bibliographystyle{mnras}
\bibliography{bibliography}

%\begin{thebibliography}{99}
%\end{thebibliography}

%%%%%%%%%%%%%%%%%%%%%%%%%%%%%%%%%%%%%%%%%%%%%%%%%%

%%%%%%%%%%%%%%%%% APPENDICES %%%%%%%%%%%%%%%%%%%%%

\onecolumn
\appendix

\section{Derivation of 3D velocity equation}\label{app:3D}

Here, we show the derivation of the velocity equation~(\ref{eq:3D_vel_id}) for ideal gas EOS (equation~\ref{eq:id}) that is based on the set of radiation hydrodynamic equations~(\ref{eq:3D}).  We also show the velocity equation~(\ref{eq:3D_vel}) for a general EOS.  Performing the scalar product of $\boldsymbol{v}$ and the momentum equation~(\ref{eq:3D_2}), using the continuity equation~(\ref{eq:3D_1}), and dividing the equation by $\rho$ yields
\begin{equation}\label{eq:A_mom}
    \boldsymbol{v} \boldsymbol{\cdot} \boldsymbol{\nabla} \lp \frac{1}{2} \left| \boldsymbol{v} \right|^2 \rp + \frac{P_{\rm gas}}{\rho} \boldsymbol{v} \boldsymbol{\cdot} \boldsymbol{\nabla} \ln P_{\rm gas} + \frac{\lambda}{\rho} \boldsymbol{v} \boldsymbol{\cdot} \boldsymbol{\nabla} E_{\rm rad} = - \boldsymbol{v} \boldsymbol{\cdot} \boldsymbol{\nabla} \phi_{\rm R}.
\end{equation}
Taking the energy equation~(\ref{eq:3D_3}), using the continuity equation~(\ref{eq:3D_1}), the definition of $\epsilon^*$, see equation~(\ref{eq:epsilon_star}), dividing the equation by $\rho$ and subtracting equation~(\ref{eq:A_mom}) gives 
\begin{equation}\label{eq:A_en}
    \epsilon_{\rm gas} \boldsymbol{v} \boldsymbol{\cdot} \boldsymbol{\nabla} \ln \epsilon_{\rm gas} - \frac{P_{\rm gas}}{\rho} \boldsymbol{v} \boldsymbol{\cdot} \boldsymbol{\nabla} \ln \rho = - c \kappa_{\rm P} \lp a T^4 - E_{\rm rad}^{(0)}\rp.
\end{equation}
Now, let us define useful thermodynamic quantities as follows
\begin{equation}\label{eq:thermo}
        \chi_{\rho, \rm gas} \equiv \left.\frac{\partial \ln P_{\rm gas}}{\partial \ln \rho}\right\vert_T, \quad \chi_{T, \rm gas} \equiv \left.\frac{\partial \ln P_{\rm gas}}{\partial \ln T}\right\vert_\rho, \quad \psi_{\rho, \rm gas} \equiv \left.\frac{\partial \ln \epsilon_{\rm gas}}{\partial \ln \rho}\right\vert_T, \quad \psi_{T, \rm gas} \equiv \left.\frac{\partial \ln \epsilon_{\rm gas}}{\partial \ln T}\right\vert_\rho.
\end{equation}
Using the above definition for the energy equation~(\ref{eq:A_en}) and rewriting it yields
\begin{equation}\label{eq:A_T}
    \boldsymbol{v} \boldsymbol{\cdot} \boldsymbol{\nabla} \ln T = \frac{1}{\psi_{T, \rm gas} \epsilon_{\rm gas}} \lp \frac{P_{\rm gas}}{\rho} - \psi_{\rho, \rm gas} \epsilon_{\rm gas}\rp \boldsymbol{v} \boldsymbol{\cdot} \boldsymbol{\nabla} \ln \rho - \frac{1}{\psi_{T, \rm gas} \epsilon_{\rm gas}} c \kappa_{\rm P} \lp a T^4 - E_{\rm rad}^{(0)}\rp.
\end{equation} 
It is useful to rewrite the continuity equation~(\ref{eq:3D_1}) into the following form
\begin{equation}\label{eq:A_mass}
    \boldsymbol{\nabla} \boldsymbol{\cdot} \boldsymbol{v} + \boldsymbol{v} \boldsymbol{\cdot} \boldsymbol{\nabla} \ln \rho = 0.
\end{equation}
Taking the momentum equation~(\ref{eq:A_mom}), utilizing the thermodynamics definitions (\ref{eq:thermo}), plugging in equation~(\ref{eq:A_T}), and using the continuity equation~(\ref{eq:A_mass}) gives
\begin{equation}\label{eq:A_vel}
    \boldsymbol{v} \boldsymbol{\cdot} \boldsymbol{\nabla} \lp \frac{1}{2} \left| \boldsymbol{v}\right|^2\rp - \lp \chi_{\rho, \rm gas} - \chi_{T, \rm gas} \frac{\psi_{\rho, \rm gas}}{\psi_{T, \rm gas}} + \frac{\chi_{T, \rm gas}}{\psi_{T, \rm gas}} \frac{P_{\rm gas} / \rho}{\epsilon_{\rm gas}}\rp \frac{P_{\rm gas}}{\rho} \boldsymbol{\nabla} \boldsymbol{\cdot} \boldsymbol{v} = \frac{\chi_{T,\rm gas}}{\psi_{T, \rm gas}} \frac{P_{\rm gas} / \rho}{\epsilon_{\rm gas}} c \kappa_{\rm P}\lp a T^4 - E_{\rm rad}^{(0)} \rp - \frac{\lambda}{\rho} \boldsymbol{v} \boldsymbol{\cdot} \boldsymbol{\nabla} E_{\rm rad} - \boldsymbol{v} \boldsymbol{\cdot} \boldsymbol{\nabla} \phi_{\rm R}.   
\end{equation}
The first velocity term in equation~(\ref{eq:A_vel}) can be written as
\begin{equation}
    \boldsymbol{v} \boldsymbol{\cdot} \boldsymbol{\nabla} \lp \frac{1}{2} \left| \boldsymbol{v}\right|^2 \rp = \left| \boldsymbol{v}\right|^2 \boldsymbol{\nabla \cdot v} - \left| \boldsymbol{v} \right|^3 \boldsymbol{\nabla \cdot} \lp \frac{\boldsymbol{v}}{\left| \boldsymbol{v}\right|}\rp.
\end{equation}
Thus, using the definitions of $q_{\rm rad}$ and $\boldsymbol{f}_{\rm rad}$ in equations~(\ref{eq:q_rad},\ref{eq:f_rad}) gives the 3D velocity equation for a general EOS that reads (cf. \citealt{lubow1975}, eq.~10)
\begin{equation}\label{eq:3D_vel}
    \left[ \left|\boldsymbol{v}\right|^2 - \lp \chi_{\rho, \rm gas} - \chi_{T, \rm gas}\frac{\psi_{\rho,\rm gas}}{\psi_{T,\rm gas}} + \frac{\chi_{T,\rm gas}}{\psi_{T,\rm gas}} \frac{P_{\rm gas}/\rho}{\epsilon_{\rm gas}}\rp \frac{P_{\rm gas}}{\rho} \right] \boldsymbol{\nabla} \boldsymbol{\cdot} \boldsymbol{v} = \left| \boldsymbol{v} \right|^3 \boldsymbol{\nabla} \boldsymbol{\cdot} \lp \frac{\boldsymbol{v}}{\left| \boldsymbol{v}\right|}\rp - \frac{\chi_{T,\rm gas}}{\psi_{T,\rm gas}} \frac{P_{\rm gas}/\rho}{\epsilon_{\rm gas}} \left| \boldsymbol{v} \right| q_{\rm rad} + \boldsymbol{v} \boldsymbol{\cdot} \boldsymbol{f}_{\rm rad} - \boldsymbol{v} \boldsymbol{\cdot} \boldsymbol{\nabla} \phi_{\rm R}.
\end{equation}
If we use ideal gas EOS (equation~\ref{eq:id}), i.e. $\chi_{\rho, \rm gas} = \chi_{T, \rm gas} = \psi_{T, \rm gas} = 1$ and $\psi_{\rho, \rm gas} = 0$, and the definition of the gas adiabatic sound speed (equation~\ref{eq:c_s_gas}) then we get the velocity equation~(\ref{eq:3D_vel_id}).

\section{Derivation of 1D radiation hydrodynamic equations and 1D velocity equation}\label{app:1D}

Under the assumptions described in Sec. \ref{sec:1D}, the divergence of a general vector field $\boldsymbol{F}_{\rm vec}$ and the gradient of a general scalar field $f_{\rm scal}$ are reduced to 1D as follows
\begin{equation}
    \boldsymbol{\nabla \cdot} \boldsymbol{F}_{\rm vec} \rightarrow \frac{1}{Q} \frac{\dd}{\dd x} \lp Q F_{\rm vec} \rp, \quad \boldsymbol{\nabla} f_{\rm scal} \rightarrow \frac{\dd f_{\rm scal}}{\dd x}.
\end{equation}
Therefore, the continuity equation~(\ref{eq:3D_1}) trivially reduces to
\begin{equation}\label{eq:B2}
    \frac{1}{Q} \frac{\dd}{\dd x} \lp \rho v Q\rp = 0.
\end{equation}
Reducing the momentum equation~(\ref{eq:3D_2}) to 1D, using the previous equation~(\ref{eq:B2}), and dividing the result by $\rho$ yields
\begin{equation}\label{eq:B3}
    v \frac{\dd v}{\dd x} + \frac{1}{\rho} \frac{\dd P_{\rm gas}}{\dd x} + \frac{\lambda}{\rho} \frac{\dd E_{\rm rad}}{\dd x} = - \frac{\dd \phi_{\rm R}}{\dd x}.
\end{equation}
Reducing equation~(\ref{eq:3D_3}) to 1D, using equation~(\ref{eq:B2}), dividing the result by $\rho v$, using the definition of the total gas specific energy $\epsilon^*$ in equation~(\ref{eq:epsilon_star}), and subtracting equation~(\ref{eq:B3}) gives
\begin{equation}\label{eq:B4}
    \frac{\dd \epsilon_{\rm gas}}{\dd x} - \frac{P_{\rm gas}}{\rho} \frac{1}{\rho} \frac{\dd \rho}{\dd x} = - \frac{c}{v} \kappa_{\rm P} \lp a T^4 - E_{\rm rad}^{(0)}\rp.
\end{equation}
Summing up equations~(\ref{eq:3D_3}) and (\ref{eq:3D_4}) and reducing the result to 1D yields a conservation law for the total luminosity being transferred through the nozzle $L_{\rm N}$ in the form $\dd L_{\rm N} / \dd x = 0$. Therefore, the luminosity $L_{\rm N} = \rm{const}.$ is given by the eqaution
\begin{equation}\label{eq:B5}
    - \frac{c \lambda}{\rho \kappa_{\rm R}} \frac{\dd E_{\rm rad}}{\dd x} = \frac{L_{\rm N}}{Q} - \lp \rho \epsilon^* + P_{\rm gas} + \frac{3-f}{2} E_{\rm rad}\rp v.
\end{equation}
We can transform equations~(\ref{eq:B2}--\ref{eq:B5}) into the set of 1D radiation hydrodynamic equations~(\ref{eq:1D}) by using simple manipulations and the definitions of $q_{\rm rad}$ and $\boldsymbol{f}_{\rm rad}$ in equations~(\ref{eq:q_rad},\ref{eq:f_rad}). 

Finally, reducing the velocity equation~(\ref{eq:3D_vel}) to 1D, canceling identical terms, and dividing it by $v$ gives the 1D velocity equation for a general EOS in the form
\begin{equation}
    \frac{1}{v}\frac{\dd v}{\dd x} = \frac{\lp \chi_{\rho, \rm gas} - \chi_{T,\rm gas}\frac{\psi_{\rho, \rm gas}}{\psi_{T, \rm gas}} + \frac{\chi_{T,\rm gas}}{\psi_{T,\rm gas}}\frac{P_{\rm gas} / \rho}{\epsilon_{\rm gas}} \rp \frac{P_{\rm gas}}{\rho} \frac{1}{Q} \frac{\dd Q}{\dd x} - \frac{\chi_{T,\rm gas}}{\psi_{T,\rm gas}}\frac{P_{\rm gas} / \rho}{\epsilon_{\rm gas}} q_{\rm rad} + f_{\rm rad} - \frac{\dd \phi_{\rm R}}{\dd x}}{v^2 - \lp \chi_{\rho, \rm gas} - \chi_{T,\rm gas}\frac{\psi_{\rho, \rm gas}}{\psi_{T, \rm gas}} + \frac{\chi_{T,\rm gas}}{\psi_{T,\rm gas}}\frac{P_{\rm gas} / \rho}{\epsilon_{\rm gas}} \rp \frac{P_{\rm gas}}{\rho}}.
\end{equation}
If we use ideal gas EOS (equation~\ref{eq:id}), i.e. $\chi_{\rho, \rm gas} = \chi_{T, \rm gas} = \psi_{T, \rm gas} = 1$ and $\psi_{\rho, \rm gas} = 0$, and the definition of the gas adiabatic sound speed (equation~\ref{eq:c_s_gas}) then we get the velocity equation~(\ref{eq:1D_vel_id}).

\section{Our previous approach and other models}\label{app:previous_and_other}

Here, we show the 1D hydrodynamic equations governing the gas flow in our original model \citepalias{cehula2023} for convenience. The equations read
\begin{subequations}\label{eq:hydro_gen}
    \begin{align}
	& \frac{1}{v}\frac{\dd v}{\dd x} + \frac{1}{\rho Q_\rho}\frac{\dd}{\dd x} (\rho Q_\rho) = 0, \\
	& v\frac{\dd v}{\dd x} + \frac{1}{\rho Q_\rho} \frac{\dd}{\dd x} (P Q_P) = -\frac{\dd\phi_{\rm R}}{\dd x}, \\
	& \frac{\dd}{\dd x} \left( \epsilon \frac{Q_P}{Q_\rho}\right) - \frac{PQ_P}{(\rho Q_\rho)^2} \frac{\dd}{\dd x}(\rho Q_\rho) = - \frac{\dd}{\dd x} \left( c_{T, \rm gas}^2 \frac{Q_P}{Q_\rho}\right), \label{eq:hydro_gen_en}
    \end{align}
\end{subequations}
where $\epsilon$ is the specific internal energy of the medium, gas plus radiation, and the characteristic cross-sections $Q_\rho$ and $Q_P$ corresponding to the density and pressure profiles respectively are defined by the following expressions
\begin{subequations}
    \begin{align}
        \rho (x) Q_\rho &= \rho(x,0,0) Q_\rho (x) \equiv \int_{Q(x)} \rho (x,y,z) \dd Q, \\
        P(x) Q_P &= P(x,0,0) Q_P(x) \equiv \int_{Q(x)} P (x,y,z) \dd Q.
    \end{align}
\end{subequations}
It holds
\begin{equation}\label{eq:Q_frac}
    Q_\rho = \frac{2 \pi}{\sqrt{BC}} c_{T, \rm gas}^2, \quad \frac{Q_P}{Q_\rho} = \frac{\gamma}{2\gamma-1}.
\end{equation} 

The optically thin MT rate of \citet{jackson2017} is given by 
\begin{equation}\label{eq:Mdot_J}
    - \dot{M}_{\rm d} \equiv \dot{M}_{\rm J} = \dot{M}_{\rm{J,0}} \exp \lp - \frac{\phi_1-\phi_{\rm ph}}{c_{T, \rm gas}^2}\rp, \quad \dot{M}_{\rm{J,0}} = \frac{2\pi}{\sqrt{\rm e}} \frac{1}{\sqrt{BC}} c_{T, \rm gas}^3 \rho_{\rm ph},
\end{equation}
where the potential difference between the L1 point $\phi_1$ and the photosphere $\phi_{\rm ph}$ is evaluated based on the following approximation
\begin{equation}\label{eq:phi_vol}
    \phi_V (r_V) = -G \frac{M_{\rm a}}{a} \left[ 1+ \frac{1}{2\lp 1+ q\rp}\right] -G \frac{M_{\rm d}}{r_V} \left[ 1+ \frac{1}{3} \lp 1+ \frac{1}{q}\rp \lp \frac{r_v}{a}\rp^3 + \frac{4}{45} \lp 1 + \frac{5}{q} + \frac{13}{q^2}\rp \lp \frac{r_v}{a}\rp^6 \right],
\end{equation}
where the volume-equivalent radius $r_V$ is defined using a sphere with a volume $V$ enclosed by the equipotential surface on the donor's side of the L1 plane, $V (\phi_V) = 4 \pi r_V^3/ 3$. The optically thick MT rate of \citet{kolb1990} is given by
\begin{equation}\label{eq:Mdot_KR}
    -\dot{M}_{\rm d} \equiv \dot{M}_{\rm{KR}} = \dot{M}_{\rm{J,0}} +  \frac{2\pi}{\sqrt{BC}} \int_{P_{\rm ph}}^{\bar{P}(R_{\rm L})} F_3 \lp\gamma \rp c_{T, \rm gas} \dd \bar{P}, \quad \text{where} \quad F_3 \lp \gamma\rp = \gamma^{\frac{1}{2}} \lp \frac{2}{\gamma+1}\rp^{\frac{\gamma+1}{2\lp\gamma-1\rp}},
\end{equation}
and to calculate $\dot{M}_{\rm J}$ and $\dot{M}_{\rm KR}$ an ideal gas EOS is used to evaluate $c_{T, \rm gas}$, see equation~(\ref{eq:id}).

\section{$1\:\rm{M}_\odot$ donor on the red giant branch}\label{app:1Msun}

We use MESA version r21.12.1 and its \texttt{1M\_pre\_ms\_to\_wd} test suite case to compute hydrostatic profiles of a donor star with initial mass $1\:\rm{M}_\odot$ and metallicity $Z = 0.02$. We evolve the stellar model to an age of 12.3 Gyr, when we obtain a red giant with the effective temperature $T_{\rm eff} = 3.4 \times 10^3\,\rm{K}$, radius $R_{\rm d} = 89\,\rm{R}_\odot$, luminosity $L_{\rm d} = 1.0 \times 10^3\,\rm{L}_\odot$, and Eddington ratio $\Gamma_{\rm Edd}$ at the surface of $1.8 \times 10^{-4}$. 

\begin{figure}
    \centering
    \includegraphics[width=0.5\columnwidth]{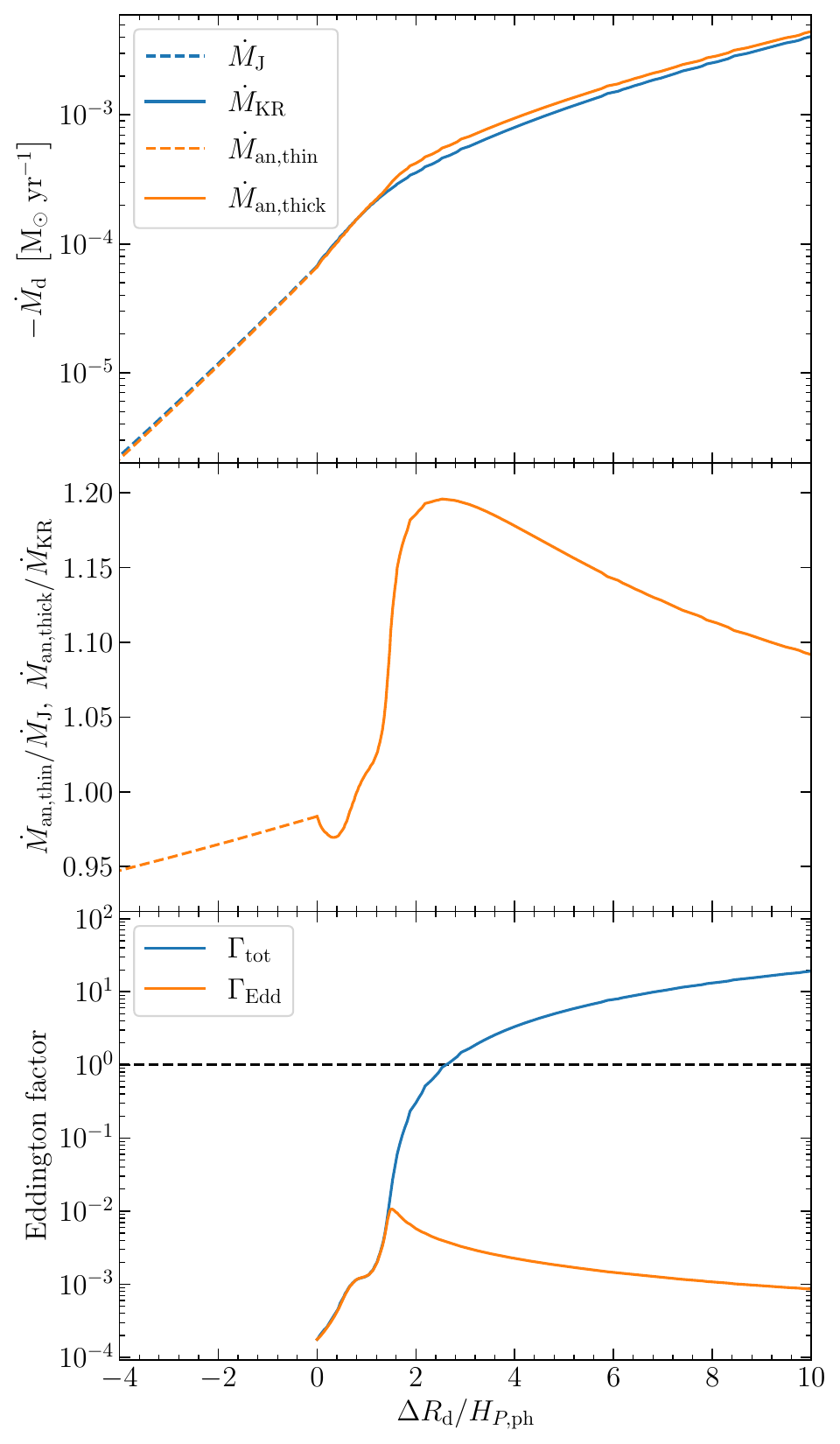}
    \caption{Comparison of our analytical MT prescription $\dot{M}_{\rm an,thin}$ and $\dot{M}_{\rm an,thick}$ with the optically thin prescription introduced by \citet{jackson2017}, $\dot{M}_{\rm J},$ and the optically thick prescription introduced by \citet{kolb1990}, $\dot{M}_{\rm KR}$.  We also show the total Eddington factor $\Gamma_{\rm tot} = L/L_\text{Edd}$ and the Eddington factor $\Gamma_{\rm Edd}$ (equation~\ref{eq:Gamma_Edd}) for comparison. The case shown is for an isolated $1 \rm{M}_\odot$ solar-like star on the red giant branch.}
    \label{fig:Mdot-an-KR-1Msun}
\end{figure}

In Fig.~\ref{fig:Mdot-an-KR-1Msun}, we compare $\dot{M}_{\rm an,thin}(\Delta R_{\rm d})$ and $\dot{M}_{\rm an,thick}(\Delta R_\dd)$ with the models of \citet{jackson2017} ($\dot{M}_{\rm J}$) and \citet{kolb1990} ($\dot{M}_{\rm KR}$). The photospheric pressure scale height in this case is $H_{P,\rm ph} = 0.011 R_{\rm d} = 0.96\,\rm{R}_\odot$. We see, as expected, that for a gas pressure dominated envelope, $P_{\rm gas} \gg P_{\rm rad}$, the differences between the two models are small (to within 20 \%).

\section{Derivation of the modified photon-tiring limit}\label{app:dotM_pt,L1}

Here, we present the derivation of the photon-tiring limit modified for the MT through the L1 point, $\dot{M}_\text{tir,L1}$ (equation~\ref{eq:mdot_tirL1}). We approximate the total luminosity transferred through the nozzle around L1 by
\begin{equation}\label{eq:C1}
    L_{\rm N} = L_{\rm d} \frac{Q}{4 \pi R_{\rm d}^2},
\end{equation}
where $Q$ is the nozzle cross-section given by equation~(\ref{eq:Q}) and $4 \pi R_{\rm d}^2$ is the surface area of the donor. The quantities in the equation~(\ref{eq:Q}) needed to evaluate $Q$ can be approximated as follows (\citealt{jackson2017}, but see also \citetalias{cehula2023})
\begin{equation}
    B = (\mathcal{A}-1) \omega^2, \quad C = \mathcal{A} \omega^2, \quad \mathcal{A} = 4 + \frac{4.16}{-0.96 + q^{1/3} + q^{-1/3}}, \quad \omega^2 = \frac{G M}{a^3},
\end{equation}
where $M = M_{\rm a} + M_{\rm d}$ is the total binary mass comprising of the accretor's and the donor's mass, $q = M_{\rm d} / M_{\rm a}$ is the binary mass ratio, and $a$ is the binary separation that can be approximated by \citep{eggleton1983approximations}
\begin{equation}
    a = \frac{R_{\rm L}}{f(q)}, \quad \text{where} \quad f(q) = \frac{0.49 q^{2/3}}{0.6q^{2/3} + \ln (1+q^{1/3})}.
\end{equation}
Using these approximations we write
\begin{equation}\label{eq:C4}
    \sqrt{BC} = \sqrt{\mathcal{A} (\mathcal{A}-1)} \lp 1 + \frac{1}{q}\rp f^3(q) \frac{G M_{\rm d}}{R_{\rm L}^3} = f_1(q)\frac{G M_{\rm d}}{R_{\rm L}^3} \approx \frac{G M_{\rm d}}{R_{\rm d}^3},
\end{equation}
where in the last approximate equation we neglect $f_1(q)$ since it holds $f_1(q) \in \langle 0.5, 1.7\rangle$, for $q \in \langle 10^{-2}, 10^2 \rangle$, and use $R_{\rm L} \approx R_{\rm d}$. Putting together the approximation for $L_{\rm N}$ in equation~(\ref{eq:C1}), the definition of $Q$ in equation~(\ref{eq:Q}), but omitting the $1/(1-\Gamma_{\rm Edd})$ factor because the luminosity being transferred through the nozzle $L_{\rm N}$ cannot diverge for $\Gamma_{\rm Edd} \to 1$, and equation~(\ref{eq:C4}) we get to order of magnitude
\begin{equation}\label{eq:C5}
    L_{\rm N} \sim L_{\rm d} \frac{c_{\rm T,gas}^2}{v_{\rm esc}^2}.
\end{equation}
To approximate the potential difference between $R_{\rm L}$ and $R_{\rm SE}$ inside the donor, we can write
\begin{equation}\label{eq:C6}
    %\bar{\phi}(R_{\rm L}) - \bar{\phi} (R_{\rm SE}) = - \frac{G M_{\rm d}}{R_{\rm L}} + \frac{G M_{\rm d}}{R_{\rm SE}} \approx \frac{G M_{\rm d}}{R_{\rm d}} \delta R_{\rm L},
    \bar{\phi}(R_{\rm L}) - \bar{\phi} (R_{\rm SE}) = - \frac{G M_{\rm d}}{R_{\rm L}} + \frac{G M_{\rm d}}{R_{\rm SE}} \approx \frac{1}{2} v_{\rm esc}^2 \delta R_{\rm L},
\end{equation}
where we again use $R_{\rm L} \approx R_{\rm d}$ and $\delta R_{\rm L} \equiv (R_{\rm L} - R_{\rm SE})/R_{\rm SE}$.

%%%%%%%%%%%%%%%%%%%%%%%%%%%%%%%%%%%%%%%%%%%%%%%%%%

% Don't change these lines
\bsp	% typesetting comment
\label{lastpage}
\end{document}